\documentclass[11pt,a4paper]{article}
\usepackage{jcappub}

\title{Gamma-ray bursts
as cosmological probes: $\Lambda$CDM vs. conformal gravity} 

\author[a,b,c]{Antonaldo Diaferio,}
\author[a,b,d,c]{Luisa Ostorero,}
\author[e]{and Vincenzo Cardone}

\affiliation[a]{Dipartimento di Fisica Generale ``Amedeo Avogadro'', Universit\`a degli Studi di Torino, Via P. Giuria 1, I-10125, Torino, Italy}
\affiliation[b]{Istituto Nazionale di Fisica Nucleare (INFN), Sezione di Torino, Via P. Giuria 1, I-10125, Torino, Italy}
\affiliation[c]{Harvard-Smithsonian Center for Astrophysics, 60 Garden Street, Cambridge, MA 02138, USA}
\affiliation[d]{Department of Physics and Astronomy, University of Pennsylvania, 209 South 33rd Street, Philadelphia, PA 12104, USA}
\affiliation[e]{INAF - Osservatorio Astronomico di Roma, via Frascati 33, 00040, Monte Porzio Catone, Roma}

\emailAdd{diaferio@ph.unito.it}
\emailAdd{ostorero@ph.unito.it}
\emailAdd{winnyenodrac@gmail.com}

\abstract{
$\Lambda$CDM, for the currently preferred cosmological
density $\Omega_0$ and cosmological constant $\Omega_\Lambda$,
predicts that the Universe expansion decelerates from early times to redshift $z\approx 0.9$
and accelerates at later times.
On the contrary, the cosmological model based on conformal gravity 
predicts that the cosmic expansion has always been accelerating. 
To distinguish between these two very different cosmologies, 
we resort to gamma-ray bursts (GRBs), which have been suggested 
to probe the Universe expansion history at $z>1$, where identified type Ia supernovae (SNe) are
rare. We use the full Bayesian approach to infer the cosmological parameters and the additional
parameters required to describe the GRB data available in the
literature. For the first time, we use GRBs as cosmological probes 
without any prior information from other data. In addition, 
when we combine the GRB samples with SNe, 
our approach neatly avoids all the inconsistencies
of most numerous previous methods that are plagued by the
so-called circularity problem. In fact, when analyzed properly, current data
are consistent with distance moduli of GRBs and SNe that can respectively be, in
a variant of conformal gravity, $\sim 15$ and $\sim 3$
magnitudes fainter than in $\Lambda$CDM. Our results indicate that the currently available
SN and GRB samples are accommodated equally well by both $\Lambda$CDM and conformal gravity 
and do not exclude a continuous accelerated expansion.
We conclude that GRBs are currently far from being effective cosmological
probes, as they are unable to distinguish between these two very different expansion histories.
}


\begin{document}
\maketitle
\flushbottom

\section{Introduction}
\label{sec:intro}

The evidence of the accelerated expansion of the Universe 
coming from extensive surveys of high-redshift type Ia supernovae (SNe)
in the late 1990s \citep{riess98,perl99}  can be accurately described by
a non-zero cosmological constant $\Lambda$ in Friedmann equations. 
For reasons that might not be fully convincing \citep{bianchi10a, kolb10, bianchi10b}, this
simple solution does not satisfy a large fraction of the scientific community.
Thus, more sophisticated models have been proposed: from arbitrary modifications
of the Einstein-Hilbert action in $f(R)$ models 
to brane-world cosmologies with extra spatial dimensions
in addition to the standard four dimensional space-time, 
to mention a few \citep[see, e.g.,][for reviews]{amendola10, nojiri10, capozziello10}. 
Despite their wildly different starting point, all these models
were conceived to describe the SN data and reproduce
the expansion history of a standard $\Lambda$ Cold Dark Matter ($\Lambda$CDM) universe. Therefore, 
these models are expected to have a decelerated period
followed by the present accelerated phase. In the $\Lambda$CDM model, the transition
occurs at redsfhit $1+z = (2\Omega_\Lambda/\Omega_0)^{1/3}$,
where $\Omega_0$ is the present average mass density and $\Omega_\Lambda$ the energy
density associated to the cosmological constant. 
According to recent measures of these parameters \citep[e.g.,][]{dunkley09}, this
transition redshift is $z\approx 0.9$. 

Before high-$z$ SN surveys were performed, the number of cosmological
models alternative to the standard Friedman model was very limited.
Notably, the steady-state cosmology required an accelerated expansion
more than forty years earlier \citep{hoyle56}. 
Before the detection of the accelerated expansion, Mannheim \cite{mann90} also proposed
conformal gravity as a cosmological model, without a cosmological constant, 
that was alternative to the standard de Sitter solution.
After the high-$z$ SN observations, Mannheim \cite{mann01,mann03} showed how conformal gravity
can easily describe these data. 
Therefore, unlike recent alternative cosmologies, the steady-state model
and conformal gravity have the remarkable property that the accelerated expansion 
is a natural feature of the model and not a requirement
for building the theory. 

Conformal gravity requires that
the expansion of the Universe has always been accelerating
and a deceleration phase never occurred. 
Thanks to this prediction, substantially different from $\Lambda$CDM,
conformal gravity is an ideal candidate
to test whether the claimed transition between the 
decelerated and accelerated expansion expected in $\Lambda$CDM is robust. 
A potentially clean and powerful test to verify the existence of a decelerating
phase in the early Universe is to extend the Hubble
diagram of SNe to very high redshifts. Unfortunately,
SNe are rarely observed at $z$ larger than $1$, and gamma-ray bursts
(GRBs) were instead proposed as possible
candidates to probe this high-redshift regime.
The first sufficiently accurate estimates of the celestial coordinates
of GRBs provided by BeppoSAX \citep{costa97} enabled the measure of the host galaxy
redshift and proved the extragalactic origin of GRBs. 
This breakthrough immediately prompted the discussion of how 
we can use GRBs to constrain the cosmological model: for example,
Cohen and Piran \cite{cohen97} discussed how GRBs with known
redshift could be used to estimate $\Omega_0$.
Atteia \cite{atteia97} also suggested to use GRBs as standard candles
based on the intrinsic dispersion of various measures
of the GRB brightness. 

A crucial step forward was the
discovery that the luminosity of GRBs appears to be correlated with
some of their temporal and spectral properties that can be directly observed 
on Earth \citep[e.g.,][]{stern99,norris00,reichart01}. 
These correlations are not yet fully understood
from first physical principles. However, their existence 
has naturally suggested the use of GRBs 
as distance indicators \citep{schaefer03,schaefer07}. GRBs have been observed
out to redshift $z=8.2$ \citep{salvaterra09,tanvir09} and are expected to be
observed to even larger redshifts in the future \citep{grindlay10,campana10}. They could thus
be a powerful cosmological probe of the
early history of the Universe expansion 
(see, e.g., \cite{ghirlanda06} for an early review). 

The obvious procedure to use GRBs as distance indicators
has two steps: (1) calibrate the correlations, and (2) 
use the calibrated correlations to estimate the
GRB luminosity distances of a given sample. We can thus 
build the GRB Hubble diagram and eventually constrain the cosmological model.
This procedure has an immediate drawback. Unlike SNe, there are no observed nearby
GRBs and the calibration of a given correlation with a GRB sample
requires the assumption of a cosmological 
model to estimate the GRB luminosities 
from their fluxes and redshifts. In principle,
within the framework of a given cosmological model,  
these correlations might be used to infer 
the luminosities, and hence the luminosity distances, of other GRBs
for which the remaining variable of the correlation is known. 
However, using these luminosity distances to 
constrain a cosmological model different from the model
used to calibrate the correlations clearly poses a problem of consistency.

A more appropriate approach is to 
extract, at the same time, the correlation coefficients and the
cosmological parameters of the model from 
the observed quantities.
To accomplish this task, it is sufficient to lay out the problem within a full Bayesian context 
and use Markov Chain Monte Carlo (MCMC) simulations 
to compute, simultaneously, 
the full probability density functions (PDFs) 
of all the parameters of interest.
This approach does not require any prior 
information on the cosmological model and yields results that are not plagued by any of the various limitations 
found in the literature \citep[e.g.,][]{friedman05, firmani05, liang05, liang06, liang08, basilakos08,
cardone09, demianski10}. In addition, we are able 
to test cosmological models both with GRBs alone, without any priors from
other probes, and by combining GRBs with other observables.

Here we apply the Bayesian analysis to the GRB sample, the SN sample, and 
the GRB and SN samples combined (sect. \ref{sec:bayesAnalysis}), 
after reviewing the basic ingredients of the conformal gravity
cosmological model in sect. \ref{sec:CG}.
We discuss our results in sect. \ref{sec:realdisc} and conclude in sect. \ref{sec:disc}.

\section{Cosmology in Conformal Gravity}
\label{sec:CG}

We investigate two variants of conformal gravity: the model proposed by Mannheim \cite{mann90},
that we will indicate with CG, 
and the kinematic conformal gravity (KCG) proposed by Varieschi \cite{varieschi09a}.

CG, which is a simplified version of the original Weyl's theory \citep{weyl18,weyl19,weyl20}, 
describes the rotation curves of disk galaxies without
resorting to dark matter \citep{mann93,mann97,mann10b}. 
In addition, CG is appealing because it appears
to be a renormalizable theory of gravitation \citep{mann09,mann10} and it is therefore suggestive of
a possible route towards the unification of the fundamental forces; moreover,
unlike other fourth-order derivative theories, CG does not suffer 
from the presence of ghosts \citep{bender08}.
However, as we mentioned above, in CG the universe has been accelerating
at all times and never went through a deceleration phase.
This very fact implies that the primordial nucleosynthesis lasts
for a more extended time interval than in the standard
model and the expected abundance of primordial deuterium is orders
of magnitudes smaller than observed \citep{knox93,eli94}.
This shortcoming of CG can be bypassed if 
astrophysical processes, rather than cosmological nucleosynthesis, supply the abundance of deuterium 
currently observed: the question is open, because the investigation
of the efficiency of all the possible mechanisms producing deuterium is still incomplete \citep{jedamzik02}. 

KCG derives from the kinematical application of the conformal symmetry to the Universe
and implies a physical interpretation of the cosmological observables radically different 
from the standard model, as we will see below. 
In principle, similarly to CG, KCG does not require the presence of dark matter,
dark energy, and an inflationary phase. In addition, KCG naturally
explains the anomalous Pioneer acceleration \citep{varieschi10}.  

We describe the basic properties of the CG and KCG cosmological models below.

\subsection{Conformal Gravity (CG)}

In CG\footnote{We use natural units with $\hbar=c=1$.} 
the field action is \citep{mann90}
\begin{equation}
I_W = -\alpha\int {\rm d}^4x\sqrt{-g}
C_{\mu\nu\kappa\lambda}C^{\mu\nu\kappa\lambda}
\end{equation}
where $C_{\mu\nu\kappa\lambda}$ is
the Weyl tensor, $\alpha$ is a coupling constant, and $g$ is the determinant
of the metric tensor $g_{\mu\nu}$.

The field equations are 
\begin{equation}
4\alpha W^{\mu\nu}= T^{\mu\nu}
\label{eq:field}
\end{equation}
with $T^{\mu\nu}$ the energy-momentum tensor and 
\begin{eqnarray}
W^{\mu\nu} &=& -\frac{1}{6}g^{\mu\nu} R^{;\lambda}_{;\lambda} + \frac{2}{3}R^{;\mu;\nu} 
 +R^{\mu\nu;\lambda}_{;\lambda}  + \cr
& \phantom{=} & - R^{\nu\lambda;\mu}_{;\lambda} - R^{\mu\lambda;\nu}_{;\lambda} 
  +\frac{2}{3} RR^{\mu\nu} + \cr
& \phantom{=} &- 2 R^{\mu\lambda}R^\nu_\lambda + \frac{1}{2} g^{\mu\nu} R_{\lambda\kappa}
R^{\lambda\kappa} -\frac{1}{6} g^{\mu\nu}R^2
\end{eqnarray}
where $R^{\mu\nu}$ and $R$ are the Ricci tensor and scalar, respectively.

The matter action can take the form \citep{mann92, mann06}
\begin{eqnarray}
I_M & =& -\int {\rm d}^4x\sqrt{-g} \left\{\frac{1}{2}S^{;\mu}S_{;\mu} - \frac{1}{12}S^2 R + \right. \cr
& \phantom{=} & +\left. \lambda S^4 + i\bar\psi\gamma^\mu[\partial_\mu +\Gamma_\mu]\psi - hS\bar\psi\psi\right\}
\end{eqnarray}
where $S(x)$ is a scalar field introduced to spontaneously break the conformal symmetry, 
$\psi(x)$ is a fermion field representing all matter, $\Gamma_\mu(x)$ is the
fermion spin connection, $\gamma^\mu(x)$ are the general relativistic
Dirac matrices, and $h$ and $\lambda$ are dimensionless coupling constants.
The above action can be extended to include more than one scalar field \citep{mann01}.  
The first three terms in the action are generated with an effective Ginzburg-Landau
theory where $S(x)$ is a phase transition condensate order parameter. Specifically,
the term $\lambda S^4$ represents the negative minimum of the Ginzburg-Landau
potential, namely the vacuum energy density \citep{mann01}. It follows that we must have
$\lambda<0$.\footnote{We note that this sign is opposite to the 
requirement $\lambda>0$ of Mannheim \cite{mann92} and Elizondo and Yepes \cite{eli94}.}

With this action and its matter and scalar field equations, the generic energy-momentum tensor is
\begin{eqnarray}
T^{\mu\nu}&=&i\bar\psi\gamma^\mu[\partial^\nu +\Gamma^\nu]\psi + \frac{2}{3}S^{;\mu}S^{;\nu} 
-\frac{1}{6} g^{\mu\nu}S^{;\kappa}S_{;\kappa} \cr 
&\phantom{=} & -\frac{1}{3} SS^{;\mu;\nu} + 
\frac{1}{3} g^{\mu\nu}SS^{;\kappa}_{;\kappa} \cr 
&\phantom{=} & -\frac{1}{6} S^2\left(R^{\mu\nu} -\frac{1}{2}g^{\mu\nu} R\right)
- g^{\mu\nu}\lambda S^4 \; .
\end{eqnarray}

Using local conformal invariance, we can set 
$S=S_0={\rm const}$ and the energy-momentum tensor becomes 
\begin{equation}
T^{\mu\nu}=T^{\mu\nu}_{\rm kin} -\frac{1}{6}S_0^2\left(R^{\mu\nu} -\frac{1}{2}
g^{\mu\nu} R\right) - g^{\mu\nu}\lambda S_0^4
\end{equation}
where $T^{\mu\nu}_{\rm kin}=i\bar\psi\gamma^\mu[\partial^\nu +\Gamma^\nu]\psi$.

Moreover, in the Friedmann-Robertson-Walker (FRW) metric
\begin{equation}
{\rm d}s^2 = -{\rm d}t^2 + a^2(t)\left[\frac{{\rm d}r^2}{1-kr^2}+r^2({\rm d}\theta^2
+\sin^2\theta {\rm d}\phi^2)\right] \; ,
\end{equation}
$W^{\mu\nu}=0$ and equation (\ref{eq:field}) implies $T^{\mu\nu}=0$,
namely 
\begin{equation}
\frac{S_0^2}{6}\left(R^{\mu\nu}-\frac{1}{2} g^{\mu\nu} R\right) = T^{\mu\nu}_{\rm kin}
-g^{\mu\nu}\lambda S_0^4 \; . 
\label{eq:einst-field}
\end{equation}
For an energy-momentum tensor of a perfect fluid, 
$T^{\mu\nu}_{\rm kin}=(\rho+p)u^\mu u^\nu + pg^{\mu\nu}$,
equation (\ref{eq:einst-field}) is similar to Einstein cosmic equations recast in the form\footnote{We use the Weinberg \cite{weinberg72} sign convention.}
\begin{equation}
-\frac{1}{8\pi G}\left(R^{\mu\nu}-\frac{1}{2} g^{\mu\nu} R\right) = T^{\mu\nu}
-g^{\mu\nu}\Lambda^\prime 
\label{eq:einst-EH}
\end{equation}
where the usual cosmological constant $\Lambda$=$8\pi G\Lambda^\prime$.
It follows that, in conformal cosmology, $G$ is replaced by the negative quantity $-3/(4\pi S_0^2)$
and the usual cosmological constant $\Lambda=8\pi G\Lambda^\prime$
by the quantity $-6\lambda S_0^2$, where $\Lambda^\prime=\lambda S_0^4$.

Equation (\ref{eq:einst-field}) and a perfect fluid energy-momentum tensor yield the differential equation for the scale factor $a$
\begin{equation}
\dot a^2 + k = -\frac{2\rho a^2}{S_0^2} - 2\lambda S_0^2 a^2
\label{eq:fried}
\end{equation}
which is identical to the standard Friedmann equation ($\dot a^2 + k = 8\pi G\rho a^2/3 + \Lambda a^2/3$) with the proper substitutions 
$G=-3/(4\pi S_0^2)$ and $\Lambda=-6\lambda S_0^2$ mentioned above. These substitutions make the
fundamental difference with the standard Friedmann models, because in CG 
both $G$ and $\Lambda^\prime$ are negative rather than positive. 
Moreover, $G$ and $\Lambda^\prime$ depends on the same parameter $S_0^2$, and they 
depend on it in the opposite way: the lower is the gravitational
cosmic repulsion (because $G$ is negative), the larger is the absolute value
of the vacuum energy density $\Lambda^\prime$.
Finally, CG is able to describe
the flat rotation curves of spiral galaxies with the aid of 
a universal constant $\gamma_0$, which is related to the geometric parameter $k$ of 
the FRW metric by the relation $\gamma_0^2=-4k$ \citep{mann97}. Therefore,
we need to have $k<0$.

In equation (\ref{eq:fried}), we can separate the matter/energy density
into a relativistic and a non-relativistic component: 
$\rho=\rho_{\rm nr}+\rho_{\rm r}=\rho_{\rm nr0}(a_0/a)^3+\rho_{\rm r0}
(a_0/a)^4$; the dependence on $a$ derives from the conservation of the energy-momentum 
tensor of a perfect fluid with an equation of state $p=w\rho$, with $w=0$ and $w=1/3$
for the non-relativistic and relativistic components, respectively; 
in the FRW metric, this equation of state implies $\rho a^{3(w+1)}={\rm const}$. 

We define 
\begin{eqnarray}
\Theta_{\rm m} &\equiv &\frac{2\rho}{H^2S_0^2}= \Theta_{\rm nr}+ \Theta_{\rm r}\equiv\frac{2\rho_{\rm nr}}{ H^2S_0^2} + \frac{2\rho_{\rm r}}{ H^2S_0^2} \; , \cr 
\Theta_\Lambda&\equiv&-\frac{2\lambda S_0^2}{ H^2} \; , \cr
\Theta_k&\equiv&-\frac{k}{ H^2 a^2}\; , 
\end{eqnarray}
where $H=\dot a/a$.
All these parameters are positive
because $\lambda<0$ and $k<0$.  Equation (\ref{eq:fried}) yields
\begin{equation}
\Theta_\Lambda+\Theta_k-\Theta_{\rm nr}-\Theta_{\rm r}=1 \; .
\end{equation}
Unlike the standard cosmology, the expansion of the scale factor in the conformal
universe accelerates at all times. In fact, 
by taking the derivative of equation (\ref{eq:fried}), we get
\begin{equation}
\ddot a = H^2a\left(\frac{\Theta_{\rm nr}}{ 2}+\Theta_{\rm r} + \Theta_\Lambda\right)
\end{equation}
which is always positive, whereas the deceleration parameter
\begin{equation}
q\equiv -\frac{\ddot a a}{ \dot a^2} = -\frac{\Theta_{\rm nr}}{ 2}-\Theta_{\rm r} - \Theta_\Lambda 
\end{equation}
is always negative.

By labelling the $\Theta$ parameters at the present time $t_0$ 
with the ``$0$'' subscript, we 
can rewrite equation (\ref{eq:fried}) as
\begin{equation}
\dot a^2a^2 = H_0^2(\Theta_{\Lambda 0}a^4 + \Theta_{k0}a^2 - \Theta_{\rm nr0}a - \Theta_{\rm r0} )\; .
\label{eq:confcosm}
\end{equation}
Again, equation (\ref{eq:confcosm}) is analogous to the standard Friedmann equation
\begin{equation}
\dot a^2a^2  = H_0^2(\Omega_{\Lambda 0}a^4 + \Omega_{k0}a^2 + \Omega_{\rm nr0}a + \Omega_{\rm r0})
\label{eq:standcosm}
\end{equation}
with the standard $\Omega_{\rm nr0,r0}=8\pi G\rho_{\rm nr0,r0}/3H_0^2$, $\Omega_{\Lambda 0}=\Lambda/3H_0^2$,
and $\Omega_{k0}=\Omega_{\rm nr0}+\Omega_{\rm r0}+\Omega_{\Lambda 0}-1$.

The main difference between the standard equation (\ref{eq:standcosm}) and its
conformal counterpart (equation \ref{eq:confcosm}) is the negative sign in front of 
the matter parameters $\Theta_{\rm nr0}$ and $\Theta_{\rm r0}$.
The left-hand side is always positive; therefore, to have real
solutions for $a(t)$, the sum of the first two terms on the 
right-hand side of equation (\ref{eq:confcosm}) must always
be larger than the sum of the latter two. On the contrary, in equation (\ref{eq:standcosm}), 
for sufficiently small $a$, the matter terms dominate and we
have the usual solutions $a\propto t^{1/2}$, when $\Omega_{\rm nr0}a$ is negligible
in the radiation dominated epoch, and $a\propto t^{2/3}$ in
the matter dominated epoch, at somewhat larger $a$'s. 

\begin{figure}
\includegraphics[angle=0,scale=.45]{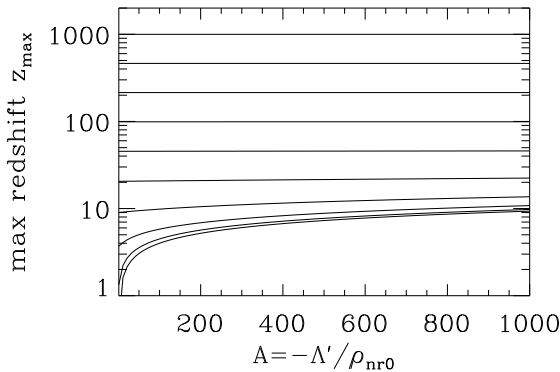}
\caption{Dependence of the maximum observable redshift $z_{\rm max}$ on 
the parameters $A$ and $B$ in equation (\ref{eq:apprxcosm}). From bottom
to top, the curves are for increasing $B=-k/(\rho_{\rm nr 0}S_0^2 a_0^2)$ in the range $1-1000$, logarithmically spaced.}
\label{fig:zmin}
\end{figure}

Therefore, in CG, $a$ never reaches the singularity $a=0$, but 
rather a lower limit $a_{\rm min}>0$ that is the root of the equation 
$\Theta_{\Lambda 0}a^4 + \Theta_{k0}a^2 - \Theta_{\rm nr0}a - \Theta_{\rm r0}=0$.
In the real Universe we can observe objects at very high redshift, $z>8$ \citep[e.g.,][]{lehnert10}.
Thus, we must have $a_{\rm min}=1/(1+z_{\rm max})<0.1$, or possibly smaller.
To obtain sufficiently small roots $a_{\rm min}$,  we can set larger and larger $\Theta_{\Lambda 0}$
and smaller and smaller $\Theta_{\rm nr0}$ and  $\Theta_{\rm r0}$. 

To understand how $a_{\rm min}$ depends on the $\Theta$ conformal
cosmological parameters, we can consider that $\rho_{\rm r0}$, 
the density of the Cosmic Microwave Background, 
is $\rho_{\rm r0}\sim 10^{-34}$~g~cm$^{-3}$ \citep{dunkley09}. 
To estimate $\rho_{\rm nr0}$ we
consider only the contribution of the luminous component
of the galaxies and assume a mass-to-light
ratio $M/L\sim 1$, because CG, in principle, does not require
the existence of any dark matter \citep{mann93}. We find $\rho_{\rm nr0}\sim 10^{-32}$~g~cm$^{-3}$.
Therefore $\Theta_{\rm r0}/\Theta_{\rm nr0}=\rho_{\rm r0}/\rho_{\rm nr0}\sim 10^{-2}$ and we can neglect
the last term in equation (\ref{eq:confcosm}), which becomes 
\begin{equation}
\dot a^2 a=H_0^2\Theta_{\rm nr0}\left(Aa^3 + Ba - 1\right)
\label{eq:apprxcosm}
\end{equation}
where $A=\Theta_{\Lambda0}/\Theta_{\rm nr0}=-\Lambda^\prime/\rho_{\rm nr0}$, 
and $B=\Theta_{k0}/ \Theta_{\rm nr0}=-k/(\rho_{\rm nr 0}S_0^2 a_0^2)$.
With this approximation, $a_{\rm min}=[(1+Q)^{1/3}+(1-Q)^{1/3}]/(2A)^{1/3}$, where $Q=\sqrt{1+
4B^3/27A}$. When $B\to\infty$, $a_{\rm min}=1/B$, whereas when $B\to 0$, $a_{\rm min}=1/A^{1/3}$.
Figure \ref{fig:zmin} shows the dependence of $z_{\rm max}$ on $A$ and $B$. Clearly,
sufficiently large $A$ and $B$, with no specific fine tuning, 
can easily accommodate the observations of high-$z$ objects.

\begin{figure}
\includegraphics[angle=0,scale=.45]{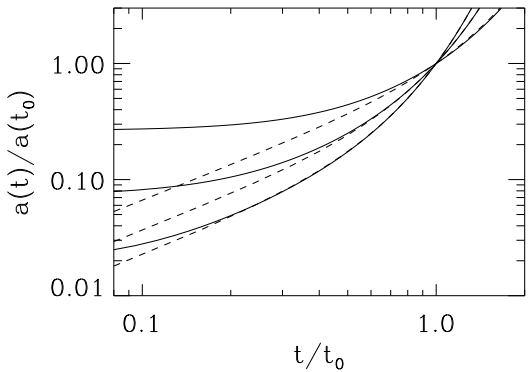}
\caption{Evolution of the scale factor $a(t)$ for different $z_{\rm max}$:
from top to bottom the solid lines are for $z_{\rm max}=2.8$, $13$, $53$.
The dashed lines show the corresponding $a(t)$ when the matter
terms $\Theta_{\rm nr0}$ and $\Theta_{\rm r0}$
are neglected. These approximated solutions have the same 
$H_0t_0$ of the exact solutions.}
\label{fig:scalefactor}
\end{figure}

For a sufficiently large scale factor $a$, we can also drop the last term in 
equation (\ref{eq:apprxcosm}) and replace
equation (\ref{eq:confcosm}) with
\begin{equation}
\dot a^2 = H_0^2(\Theta_{\Lambda 0}a^2 + \Theta_{k0}) \; .
\label{eq:acccosm}
\end{equation}
The integration of this equation trivially is 
\begin{equation}
a(t) =\sqrt\frac{\Theta_{k0}}{ \Theta_{\Lambda 0}} \sinh  y
\label{eq:accel}
\end{equation}
with $y=\sqrt{\Theta_{\Lambda 0}} H_0 t$. Clearly,
this solution is invalid when $a\to a_{\rm min}$. 
By imposing $a(t_0)=a_0=1$ for the scale factor at the present time,
we have $\sinh y_0 = \sqrt{\Theta_{\Lambda 0}/\Theta_{k0}}$ and from $q=-\ddot a a
/\dot a^2 $, we find $\Theta_{k0} = -\Theta_{\Lambda 0} (1+q_0)/q_0$.

We can determine the present age of the universe $t_0$ from the relation $H(t_0)=H_0$.
We find 
\begin{equation}
H_0 t_0 = \frac{1}{ \sqrt{-q_0}} {\rm atanh}\left( \sqrt{-q_0}\right) \; .
\label{eq:h0t0appr}
\end{equation}
Of course, $\Theta_{\Lambda 0}+\Theta_{k0}=1$, because the mass parameters are negligible
in this approximation.

Figure \ref{fig:scalefactor} shows the solution $a(t)$ of equation (\ref{eq:confcosm}) 
for three different $z_{\rm max}$ (solid lines). The dashed lines
show equation (\ref{eq:accel}) with the same $H_0 t_0$ as the exact solutions;
in other words, these approximated solutions have 
a different set of $\Theta$'s but the same slope at $t_0$ of the exact solutions.
Of course, at increasing $z_{\rm max}$ the two solutions of 
equations (\ref{eq:confcosm}) and (\ref{eq:acccosm})
start to become indistinguishable at earlier and earlier times.

In the accelerated limit, we can analytically compute the luminosity distance $d_L=x(1+z)$, where
the radial distance $x$ is implicitly defined by the relation
\begin{equation}
\int_0^x \frac{{\rm d}r}{ \sqrt{1-kr^2}} = \int_{t_1}^{t_0}\frac{ {\rm d}t}{ a(t)}
\end{equation}
and $t_1$ is the time of the light emission.
One finds
\begin{equation}
\int_{t_1}^{t_0}\frac{ {\rm d}t}{ a(t)} = \frac{1 }{ \sqrt{\Theta_{k0}}  H_0}  \ln \left[\frac{\tanh(y_0/2)}{ 
\tanh(y_1/2)}\right]
\end{equation}
whereas
\begin{equation}
\int_0^x \frac{{\rm d}r}{ \sqrt{1-kr^2}} = \frac{1}{ \sqrt{-k}} \sinh^{-1} (\sqrt{-k} x)
\end{equation}
with $k<0$.  With some algebra, we obtain 
\begin{equation}
d_L = \frac{(1+z)^2}{ q_0H_0 } \left[\left(1+q_0 -
\frac{q_0}{ (1+z)^2}\right)^{1/2}-1\right] \; .
\label{eq:lumdist}
\end{equation}

Because equation (\ref{eq:accel}) is an excellent approximation 
to the exact solution of equation (\ref{eq:confcosm}) at sufficiently
late times, we 
can safely apply equation (\ref{eq:lumdist}) to the real
Universe and consider the  
distance modulus $m-M=\mu(z; q_0)=25+5\log_{10}[d_L(z; q_0)/{\rm Mpc}]$,
where $q_0$ is the only free parameter.

\subsection{Kinematic Conformal Cosmology (KCG)}\label{sec:KCG}

The starting point of KCG \citep{varieschi09a} is the
static Schwarzschild solution of CG which yields the metric
\begin{equation}
{\rm d}s^2 = -B(r) c^2{\rm d}t^2 + \frac{{\rm d}r^2}{ B(r)} + r^2{\rm d}\Omega
\label{eq:Varmetr}
\end{equation}
where
\begin{equation}
B(r) =1 -\frac{\beta(2-3\beta \gamma)}{ r}  - 3 \beta\gamma
+ \gamma r  - \kappa r^2 \; ;
\end{equation}
$\beta$ and $\gamma$ depend on the source mass and $\kappa$ is a constant. 
If we consider sufficiently large distances from the mass source 
[$r\gg \beta(2-3\beta \gamma)$] and ignore the
term $\beta\gamma$, that rotation velocities of spiral
galaxies suggest to be negligible \citep{mann93,mann97},  $B(r)$ simplifies to 
\begin{equation}
B(r) = 1+\gamma r - \kappa r^2 \; .
\end{equation}
By using the local conformal invariance, we can show that this metric 
is conformal to the standard FRW metric
\begin{equation}
{\rm d}s^2 = -c^2 {\rm d}{\bf t}^2 + a^2({\bf t})\left({{\rm d}{\bf r}^2\over 1-{\bf k r}^2} 
+ {\bf r}^2{\rm d}\Omega\right)
\label{eq:VarmetrFRW}
\end{equation}
with ${\bf k}=k/\vert k\vert=0,\pm 1$, and $k=-\gamma^2/4 -\kappa$. We omit 
the transformations between the coordinates $(r,t)$ and
the coordinates $({\bf r},{\bf t})$ which can be found in \cite{varieschi09a, varieschi09b}.

The local conformal invariance introduces a dependence of the length and time units on
the local metric. The redshift 
\begin{equation}
1+z = {a({\bf t}_0)\over a({\bf t})}
\end{equation}
can thus be interpreted as the ratio 
between the wavelength $\lambda({\bf r},{\bf t})$ of the radiation
emitted by the atomic transitions at the time and location of the source 
and the wavelength $\lambda(0,{\bf t}_0)$ of the same atomic 
transitions measured on Earth now:
\begin{equation}
1+z = {\lambda({\bf r},{\bf t})\over \lambda(0,{\bf t}_0)} \; . 
\end{equation} 
Unlike the standard cosmology, where the measured redshift is 
due to the expansion of the scale factor $a$, in KCG, 
the redshift origins from the change of length and time units over the cosmological
time and space.

With this interpretation of the cosmic redshift, we can
derive the dependence of the scale factor $a$ on ${\bf r}$ or ${\bf t}$ without 
explicitly solving the field equations. 
In fact, $1+z=\lambda({\bf r},{\bf t})/\lambda( 0,{\bf t}_0)$
is the ratio of two frequencies $\nu( 0,{\bf t}_0)/ \nu({\bf r},{\bf t})$ which reduces
to the ratio of two time intervals, or the square-root of the ratio 
of the time-time components $g_{00}$
of the metric at the two different locations. 

\begin{figure}
\includegraphics[angle=0,scale=.45]{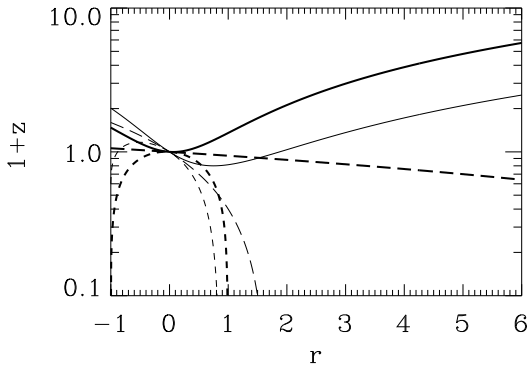}
\caption{Relation between redshift $z$ and radial coordinate ${\bf r}$ of 
the FRW metric in KCG.
The solid, long-dashed and short-dashed lines are for ${\bf k}=-1,0$, and
$1$, respectively. The bold (thin) lines are for $\delta=0.06$ ($0.6$).} 
\label{fig:Varscalefactor}
\end{figure}

With the metric (\ref{eq:Varmetr}), 
\begin{equation}
1+z = \sqrt{-g_{00}(0,t_0)\over-g_{00}(r,t)} = {1\over \sqrt{1+\gamma r-\kappa r^2}} \; ,  
\end{equation}
which yields, with the proper coordinate transformations to the metric (\ref{eq:VarmetrFRW}), 
\begin{equation}
1+z = {a(0)\over a({\bf r})} = \sqrt{1-{\bf kr}^2}-\delta {\bf r}
\label{eq:Varred}
\end{equation}
where 
\begin{displaymath}
\delta ={\gamma \over 2}\left\{\begin{array}{ll}
 \vert k \vert^{-1/2}   &  k \ne 0  \\
 1       &         k=0 \; . \end{array} \right.
\end{displaymath}
Figure \ref{fig:Varscalefactor} shows the redshift $1+z=a(0)/a({\bf r})$
as a function of the radial coordinate ${\bf r}$. The origin
of ${\bf r}$ is our observer location, the coordinate ${\bf r}$ indicates locations
of sources whose radiation has already reached us (if ${\bf r}>0$) or will reach us in the future
(if ${\bf r}<0$). Clearly, the observed redshift $z>0$ appears only
if ${\bf k}=-1$ when ${\bf r}>0$. Therefore models with ${\bf k}=0$ or $1$ are not viable.
The case ${\bf k}=-1$ also has a range of positive ${\bf r}$ where $z<0$; 
spectra of sources at these locations would be blueshifted.
However, this range of ${\bf r}$ decreases with decreasing
$\delta$ and can be easily accommodated in the local neighborhood of the Solar
System. Varieschi \cite{varieschi09b, varieschi10} actually suggests that this feature
provides a solution to the Pioneer anomaly \citep{anderson02}.\footnote{More
recent investigations seem to indicate 
that the measured anomalous acceleration of the Pioneer 10 and 11 spacecraft is not 
a gravitational effect but 
it rather is a thermal acceleration \citep[e.g.,][]{francisco11} that might not even point  
to the Sun \citep{turyshev11}.}
When ${\bf k}=-1$, equation (\ref{eq:Varred}) 
yields
\begin{equation}
{\bf r}={\delta(1+z)\pm\sqrt{(1+z)^2-(1-\delta^2)}\over 1 - \delta^2}
\label{eq:Varr}
\end{equation}
and the two locations where $z=0$ are ${\bf r}=0$ and ${\bf r}_{\rm rs}=2\delta/(1-\delta^2)$.
There is also a minimum (negative) redshift at ${\bf r}_{\rm min}=\delta/\sqrt{1-\delta^2}$;
${\bf r}_{\rm min}$ is a real number only if $\vert \delta\vert<1$.

To derive the luminosity distance $d_{\rm L}$ in KCG,
we consider the following argument. The new interpretation of redshift implies that
time (i.e. $1/\nu$) and length (i.e. $\lambda$) scale as
\begin{equation}
\Delta l_z=(1+z)\Delta l_0
\end{equation}
\begin{equation}
\Delta t_z=(1+z)\Delta t_0 \; ,
\end{equation}
where the subscript $0$ indicates units of the given quantity associated to objects
which share the same location (in space and time) of 
the observer at the origin (namely us at ${\bf r}=0$) 
and the subscript $z$ indicates the same quantities associated to objects at redshift $z\ne 0$,
measured by the same observer at the origin. For example, for  an
atomic transition that happens here we measure the frequency $\nu_0$;
for the same atomic transition that happens at redshift $z>0$ we measure (here) the
lower frequency $\nu_z=\nu_0/(1+z)$. It is important to emphasize that 
this frequency change is due to the different location in space and time
of the atom and not to the cosmic expansion as in the standard model.
The stretching of the space-time manifold also allows for 
a generic scaling of the mass:
\begin{equation}
\Delta m_z=f(1+z)\Delta m_0
\end{equation}
where $f(1+z)$ is an arbitrary function. Clearly, the energy $\Delta E\propto 
\Delta l^2 \Delta t^{-2} \Delta m$ scales with the same factor $f(1+z)$:
\begin{equation}
\Delta E_z=f(1+z)\Delta E_0 \; .
\end{equation}

According to these scaling laws, the relation between the
luminosity $L_z$ of a source at $z\ne 0$ and the luminosity $L_0$
of the same source located at $z=0$ is
\begin{equation}
L_z=L_0 {f(1+z)\over 1+z} \; .
\label{eq:LzL0}
\end{equation}
In other words, $L_z$ is the luminosity we measure
on Earth when the source is at position ${\bf r}$ where the redshift is $z\ne 0$.
In practice, we do measure the flux $F=L_z/4\pi d_L^2$, and not $L_z$. 
Classically, $L_z$ is constant and the measured flux
only depends on the distance $d_L$. In KCG, $L_z$ is not constant, but depends
on the location of the source and scales according to equation (\ref{eq:LzL0}). To account
for this scaling, Varieschi generalizes the definition of flux to $F\sim L_0/4\pi d^{a_{\rm V}}$,
where $L_0$ is now constant, because it is the luminosity of the source at $z=0$,
and the dependence of $L_z$ on $z$ enters the power $a_{\rm V}$.
Without a correction factor, the case $a_{\rm V}\ne 2$ does not of course yield
a quantity with the dimensions of a flux. Therefore, the proper generalization
suggested by Varieschi is
\begin{equation}
F(d_L)={L_0\over 4\pi d_L^2 } \left(d_{\rm rs}\over d_L\right)^{a_{\rm V}}
\label{eq:Varflux}
\end{equation} 
where $d_{\rm rs}$ is the luminosity distance of the source at
${\bf r}={\bf r}_{\rm rs}$ where $z=0$. 
Of course $F(d_L)=L_z/4\pi d_L^2$ still holds. By combining
this relation with equations (\ref{eq:Varflux}) and 
(\ref{eq:LzL0}) we determine the scaling function
\begin{equation}
{f(1+z)\over 1+z} =   \left(d_{\rm rs}\over d_L\right)^{a_{\rm V}} \; .  
\label{eq:Varf(1+z)}
\end{equation}

In standard cosmology, the energy per unit time received on Earth 
is dimmed by a factor $(1+z)^2$ because
of the redshift of the photon frequency and the time interval dilation.
This dimming originates a factor $(1+z)$ in the luminosity distance $d_L=a_0x(1+z)$,
where $a_0$ is the scale factor at the present time, and $x$ is
the radial coordinate of the FRW metric.
In KCG, neither the energy nor the time
are affected by the expansion of the scale factor. Therefore, the 
luminosity distance simply is
\begin{equation}
d_{\rm L} = a({\bf t}_0){\bf r} \; 
\label{eq:VardL}
\end{equation}
where the radial coordinate ${\mathbf r}$ of the object
at redshift $z$ (equation \ref{eq:Varr}) is evaluated by the observer at the origin,
namely with $\delta=\delta({\mathbf t}_0)\equiv\delta_0$.
We thus have $d_{\rm rs}=a({\mathbf t}_0)2\delta_0/(1-\delta_0^2)$ and
\begin{equation}
{f(1+z)\over 1+z} = 
   \left[2\delta_0 \over \delta_0(1+z)+\sqrt{(1+z)^2-(1-\delta_0^2)}\right]^{a_{\rm V}} \; .
\label{eq:Varffin}
\end{equation}
In the above equation we have used equation (\ref{eq:Varr}) with the positive sign,
which is the solution for $z>0$ and ${\bf r}>{\bf r}_{\rm rs}$.

In standard cosmology, the distance measure 
$\mu=m-M=-2.5\log_{10}[F(d_{\rm L})/F(d_{\rm ref})]=
2.5(2+a_{\rm V})\log_{10}(d_{\rm L}/d_{\rm ref})$ clearly
has $a_{\rm V}=0$ and $d_{\rm ref}$ arbitrarily chosen to be $d_{\rm ref}=10$~pc.
In KCG, $a_{\rm V}$ is unknown, and it is natural to choose
$d_{\rm ref}=d_{\rm rs}$ where $z=0$.
We thus obtain
\begin{eqnarray}
\mu(z)&=&2.5(2+a_{\rm V})\times \cr
&\times & \log_{10} \left[\delta_0(1+z)+\sqrt{(1+z)^2-(1-\delta_0^2)}\over 2\delta_0
\right] \; .
\label{eq:Varmu}
\end{eqnarray}

Obviously, this relation relies on the arbitrarily chosen equation (\ref{eq:Varflux}) 
which leads to the definition of $f(1+z)$ (equation \ref{eq:Varffin}). Therefore, although this
choice appears to provide 
a good description of the SN data, it might not necessarily be the final
correct choice \citep{varieschi09b}. 

\section{Bayesian analysis}
\label{sec:bayesAnalysis}

We now apply the Bayesian analysis (see appendix \ref{sec:bayesParam}) 
to the GRB sample to derive the cosmological parameters of $\Lambda$CDM and
of our two alternative models along with the coefficients of four GRB correlations.
We also derive the cosmological parameters by
applying the Bayesian analysis to a SN sample only
and to the combined GRB and SN samples.

\begin{figure*}
\includegraphics[angle=0,scale=.25]{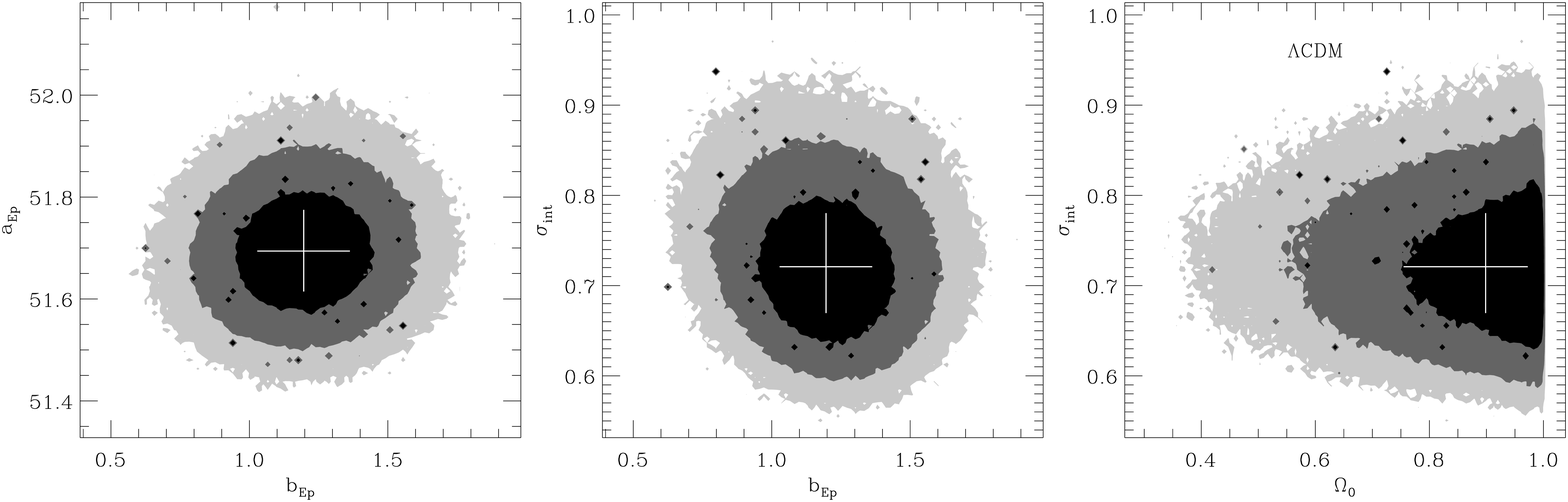}
\includegraphics[angle=0,scale=.25]{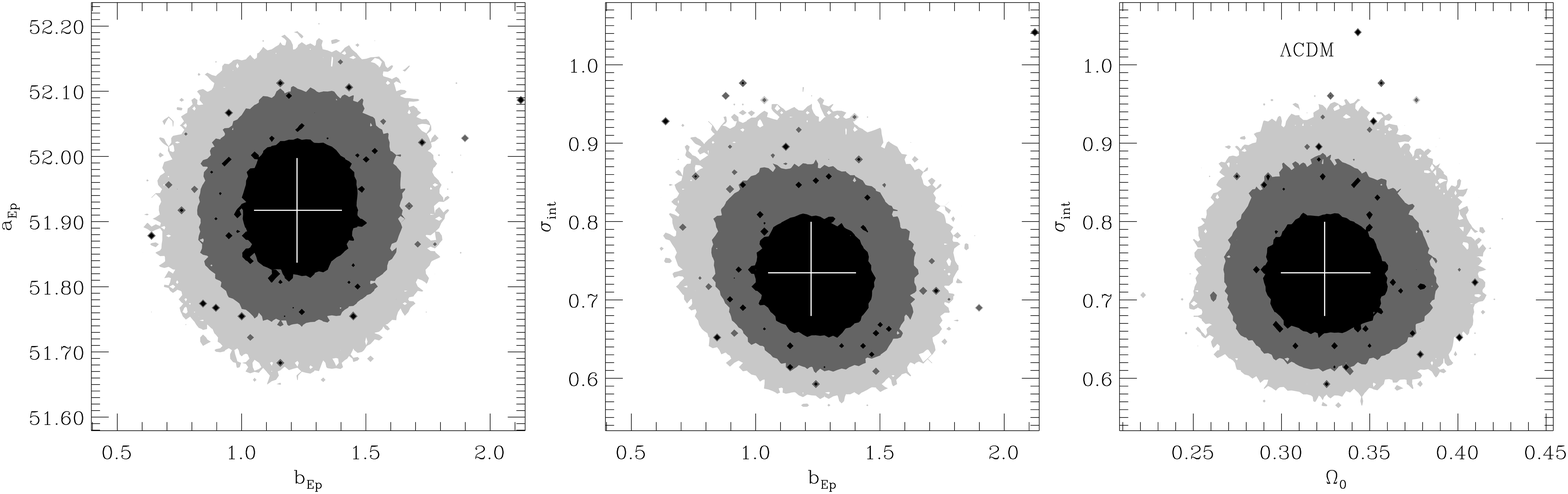}
\caption{The marginalized PDFs of $\Omega_0$ and of the parameters $a$, $b$, and $\sigma_{\rm int}$ 
of the $L-E_{\rm peak}$ GRB correlation in the $\Lambda$CDM model. 
Black, grey, and light-grey shaded regions
correspond to the 68.3, 95.4 and 99.7 percent confidence levels, respectively.
The crosses show the median values and their marginalized 1-$\sigma$ uncertainty. 
The top panels show the PDFs when the Bayesian analysis is applied to 
the GRB sample alone, the bottom panels show the PDFs when the GRB and SN samples are 
combined.}
\label{fig:GRBLCDM}
\end{figure*}

\begin{figure*}
\includegraphics[angle=0,scale=.25]{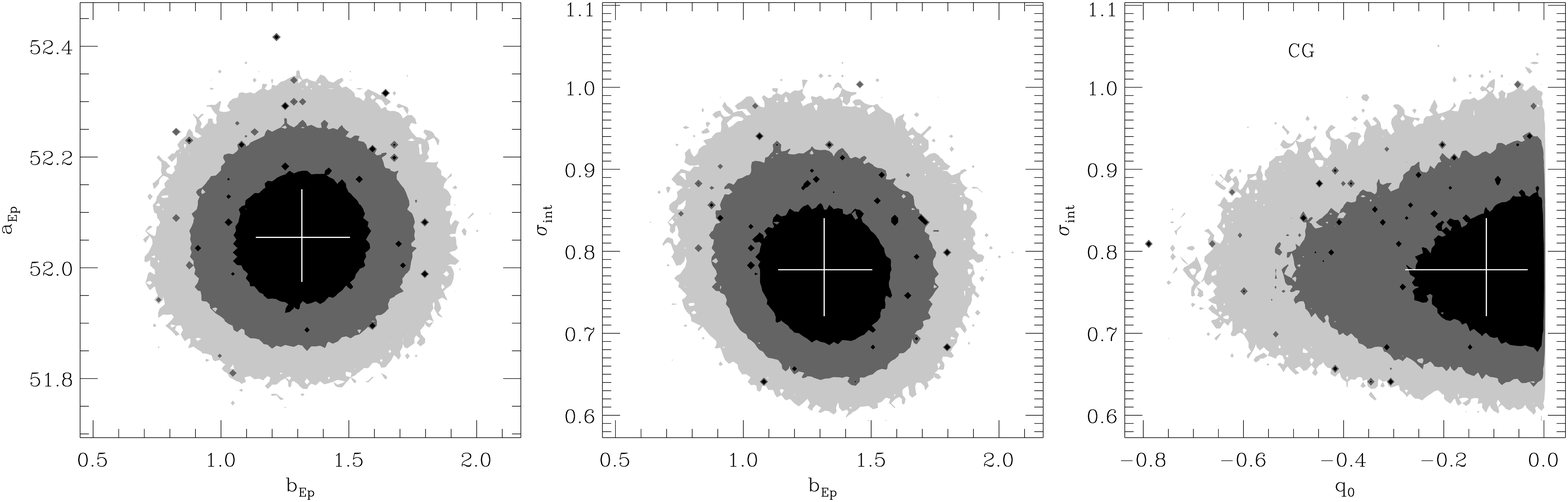}
\includegraphics[angle=0,scale=.25]{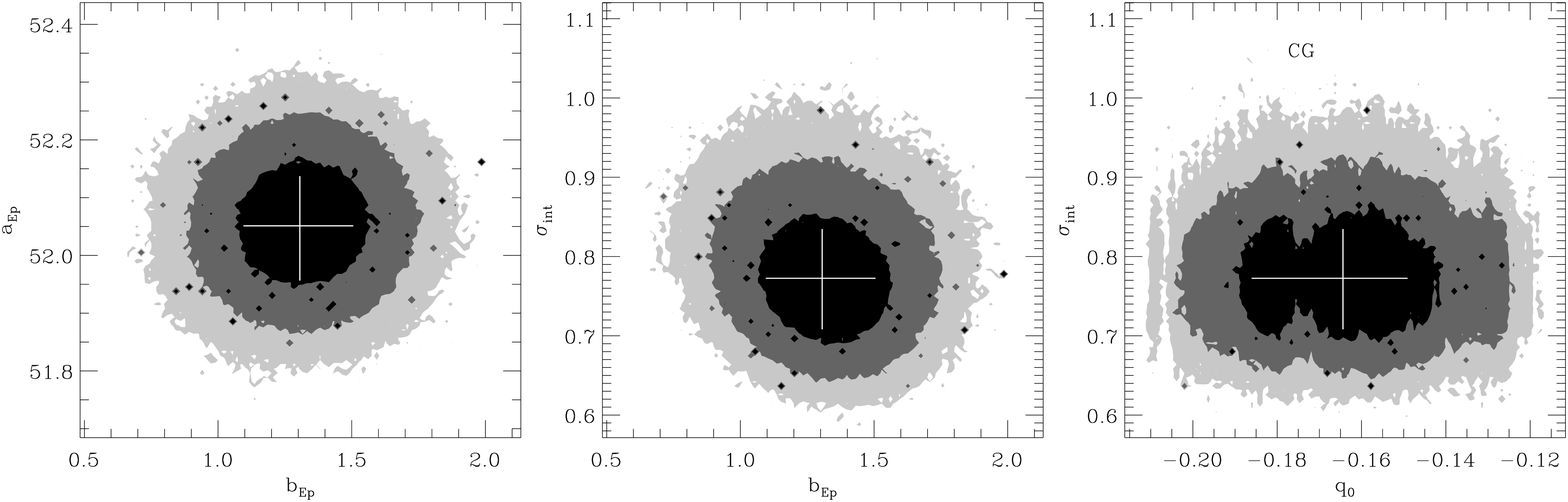}
\caption{Same as figure \ref{fig:GRBLCDM} for the CG model.}
\label{fig:GRBMann}
\end{figure*}

\begin{figure*}
\includegraphics[angle=0,scale=.25]{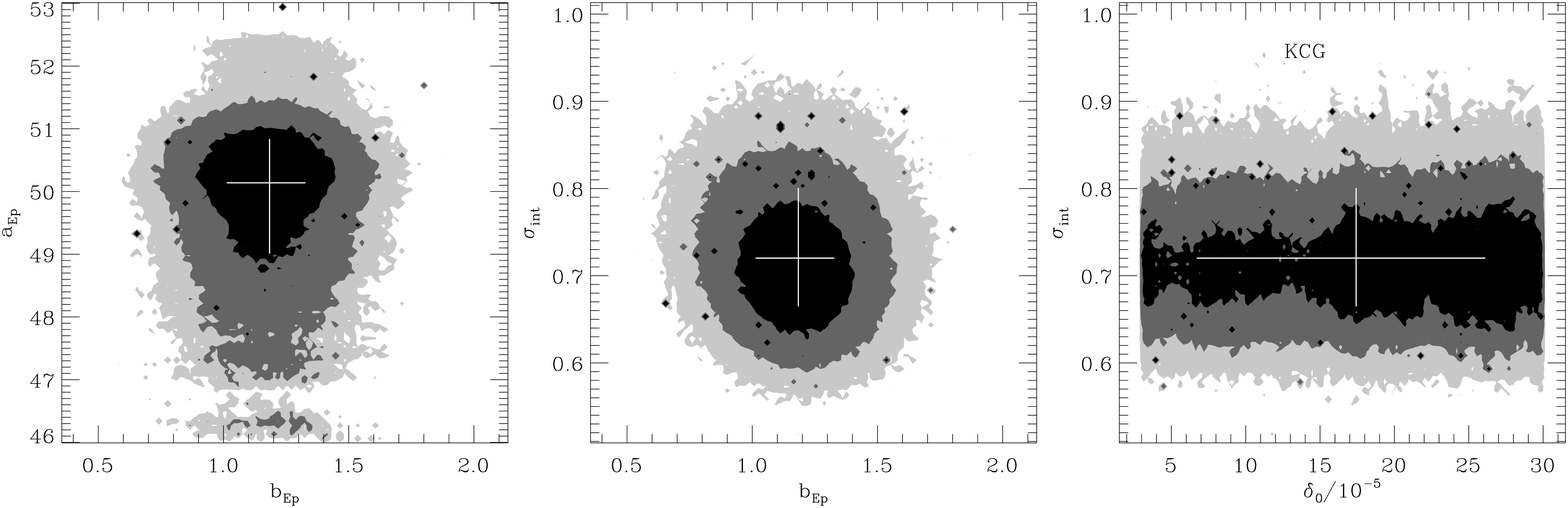}
\includegraphics[angle=0,scale=.25]{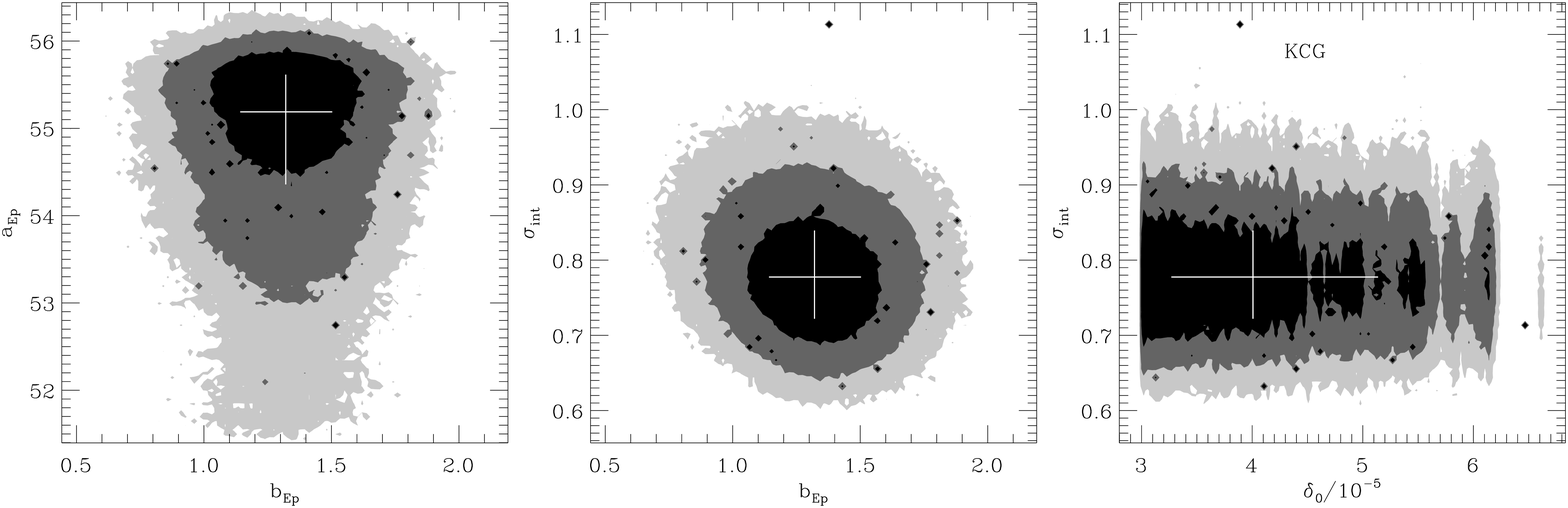}
\caption{Same as figure \ref{fig:GRBLCDM} for the KCG model.}
\label{fig:GRBVar}
\end{figure*}

\subsection{GRB sample }
\label{sec:GRBalone}

Among the distance indicators that correlate with the
GRB luminosity $L=4\pi d_L^2 P_{\rm  bol}$, 
where $P_{\rm  bol}$ is the bolometric peak flux, we consider the following \citep{schaefer07}:
\begin{enumerate} 
\renewcommand{\theenumi}{(\arabic{enumi})}
\item $\tau_{\rm lag}$, the time delay 
between the soft and the hard light curve; 
\item $\tau_{\rm RT}$, the minimum time over which the light
curve rises by half the peak flux of the pulse; 
\item $V$, the variability of the light curve defined in \cite{schaefer07};
\item $E_{\rm peak}$, the photon energy where the spectral energy distribution $\nu F_\nu$ peaks.
\end{enumerate}

As we mention in the Introduction, we do not follow the inappropriate
procedure of assuming a cosmological model, calibrating the relations,
and deriving the GRB Hubble diagram. We instead apply the Bayesian analysis
directly to the observables.  Our GRB data set is $\{P_{\rm bol}^i,\{Q^i\}_{j=1,4},z^i, {\mathbf S^i}\}$, where $\{Q^i\}_{j=1,4}=\{\tau_{\rm lag}^i,\tau_{\rm RT}^i,V^i,E_{\rm peak}^i\}$
and ${\mathbf S^i}$ is the vector of uncertainties of all the measures,
except for the redshift $z^i$.

Our task is the determination of the multi-dimensional PDF of the
parameters $\theta=\{\{a,b,\sigma_{\rm int}\}_{j=1,4},{\mathbf p}\}$, 
where $ (a_j,b_j) $ are the parameters of the four GRB correlations. In the $\Lambda$CDM and CG models
\begin{equation}
\log_{10} P_{\rm bol} = a_j + b_j \log_{10} Q_j - \log_{10}[4\pi d_L^2(z,{\mathbf p})]\; .
\end{equation}
To mimic additional hidden parameters in the relations
we assume that $\log_{10} P_{\rm bol}^i$ is a random variate with mean
\begin{equation}
\log_{10} P_{\rm bol} = a_j + b_j \log_{10} Q_j^i - \log_{10}[4\pi d_L^2(z^i,{\mathbf p})]
\end{equation}
and variance $\sigma_{\rm int}^2$ (e.g., \cite{dago05}, \cite{andreon10}); 
finally, ${\mathbf p}$ is the vector of 
the cosmological parameters:
in practice, ${\mathbf p}=\Omega_0$, with $\Omega_{\Lambda0}=1-\Omega_0$
for $\Lambda$CDM,\footnote{For the sake of simplifying the algorithm implementation, 
for the $\Lambda$CDM model, we use the analytic 
approximation to $d_L$ in flat universes provided by Pen \cite{pen99}. The formula
is 0.4 percent accurate when $\Omega_0$ is in the range $[0.2,1]$.}
and ${\mathbf p}=q_0$ for CG. For both $\Lambda$CDM and CG, the 
Hubble constant is an input parameter and we
use $H_0=73\pm 2{\rm (statistical)} \pm 4{\rm (systematic)} $~km~s$^{-1}$~Mpc$^{-1}$
\citep{freedman10}.

In KCG, the Hubble constant does not enter the estimate of $d_L$ (equations \ref{eq:Varr} and \ref{eq:VardL}). The flux measured on Earth of the source at redshift $z$,
whose luminosity measured on Earth is $L_z=L_0f(1+z)/(1+z)$, is 
\begin{equation}
P(d_L)={L_0\over 4\pi d_L^2} \left(d_{\rm rs}\over d_L\right)^{a_{\rm V}}\, , 
\end{equation}
where $d_{\rm rs}=a({\mathbf t}_0)2\delta_0/(1-\delta_0^2)$. 
Suppose the
expected relation is between quantities associated to GRBs if they
were at $z=0$; in other words, suppose we expect relations of the form 
$\log_{10} L_0 = a + b\log_{10} Q_0$. In this case, we have
\begin{eqnarray}
\log_{10} P_{\rm bol}(d_L) &=&  a_j + b_j\log_{10} Q_{0j} - \log_{10}(4\pi d_L^2) + \cr
&-& a_{\rm V} \log_{10}\left(d_L\over d_{\rm rs}\right)\; .   
\end{eqnarray}
$Q_{0j}$ is either a time 
$\tau_0=\tau_z/(1+z)$ or a frequency $E_0=E_z(1+z)$:\footnote{For $E_z$, we use the frequency scaling relation rather than the energy
relation $E_0=E_z/f(1+z)$ discussed in section \ref{sec:KCG},
because the X-ray photons originated in the GRBs of our
sample are detected via photoelectric or Compton effects, and in these
electron-photon interactions the electron is ultimately
sensitive to the photon frequency.} $\tau_z$ is the 
time of the source at redshift $z$ measured on Earth and $E_z$
is the frequency, measured on
Earth, of the photons emitted at the source at redshift $z$.
Finally, the cosmological parameters in KCG are ${\mathbf p}=[a_{\rm V}, \delta_0, a({\mathbf t}_0)]$.

\begin{figure*}
\includegraphics[angle=0,scale=0.8]{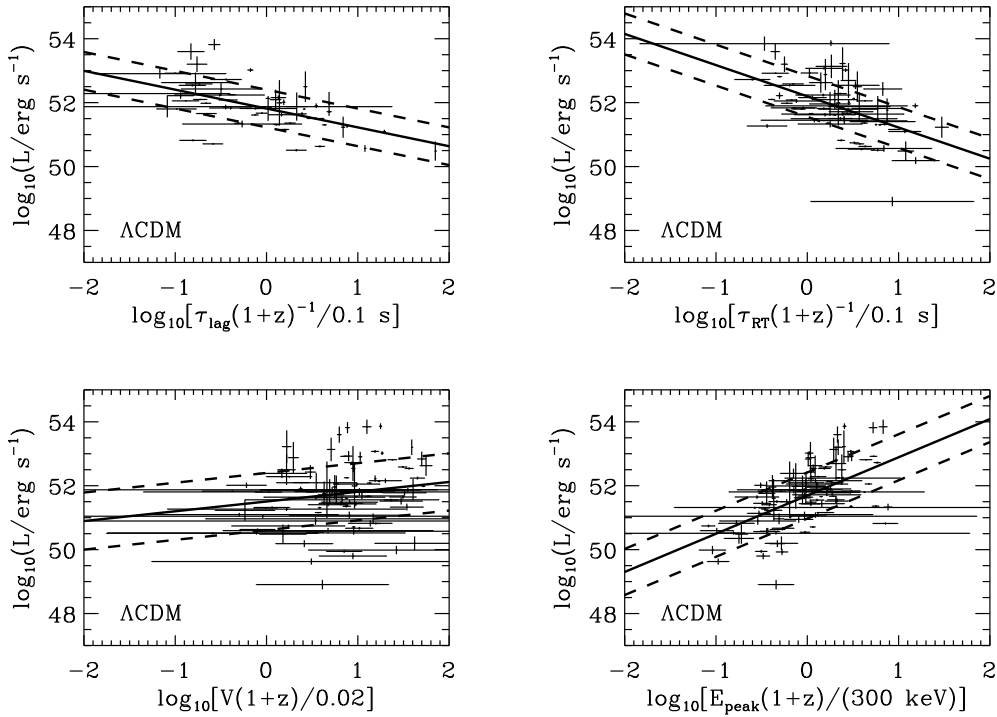}
\caption{The four GRB correlations when the Bayesian analysis is
applied to the GRB sample alone in the $\Lambda$CDM model. The crosses
show the GRB measurements with the $1$-$\sigma$ errorbars. The thick solid
straight-lines are the correlations according to the coefficients listed in Table
\ref{tab:GRBrelations}. The dashed lines show the $\pm\sigma_{\rm int}$ standard deviation of 
each relation. }
\label{fig:GRBcorrLCDM}
\end{figure*}

\begin{figure*}
\includegraphics[angle=0,scale=0.8]{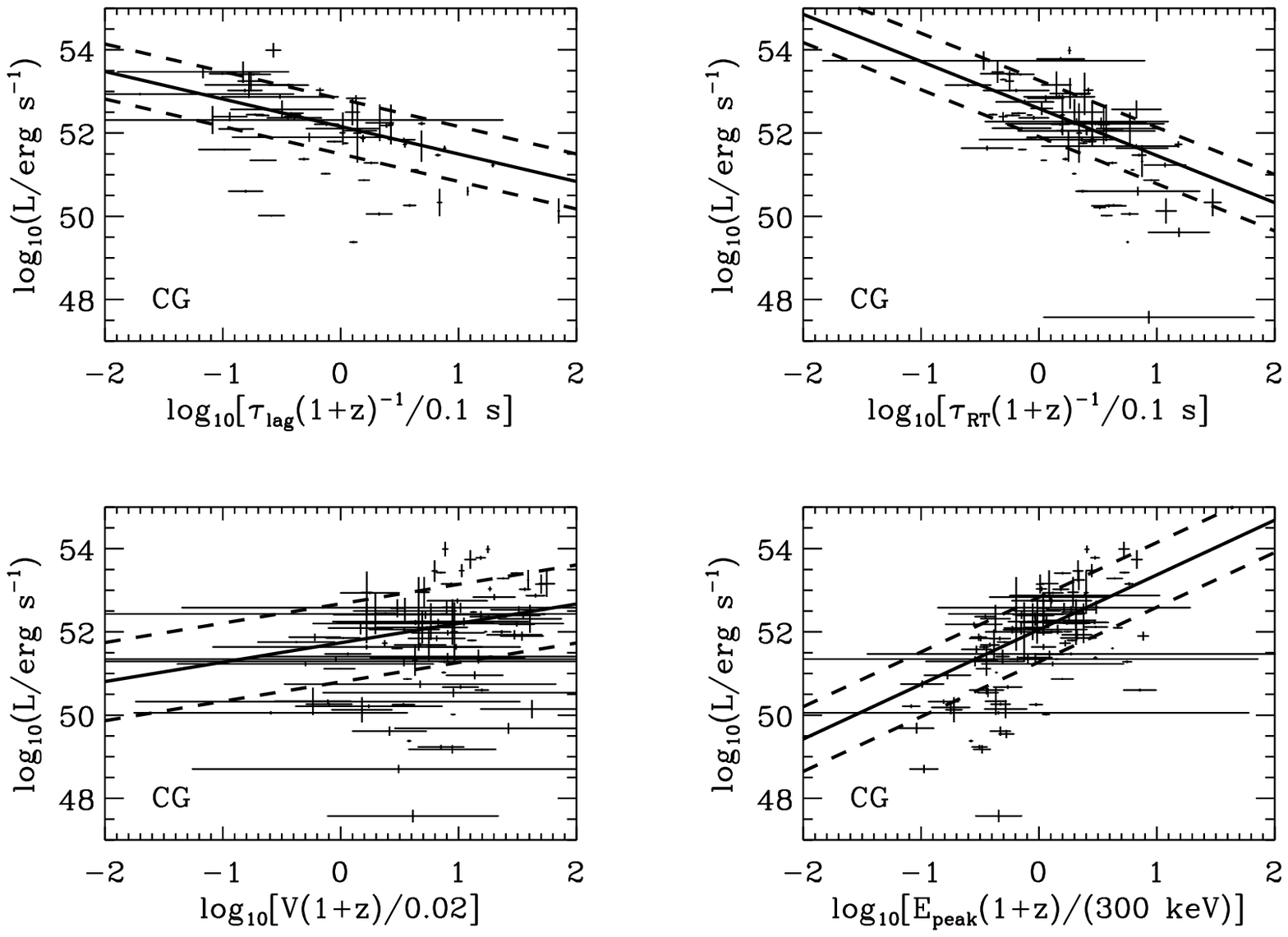}
\caption{Same as figure \ref{fig:GRBcorrLCDM} for the CG model.}
\label{fig:GRBcorrCG}
\end{figure*}

\begin{figure*}
\includegraphics[angle=0,scale=0.8]{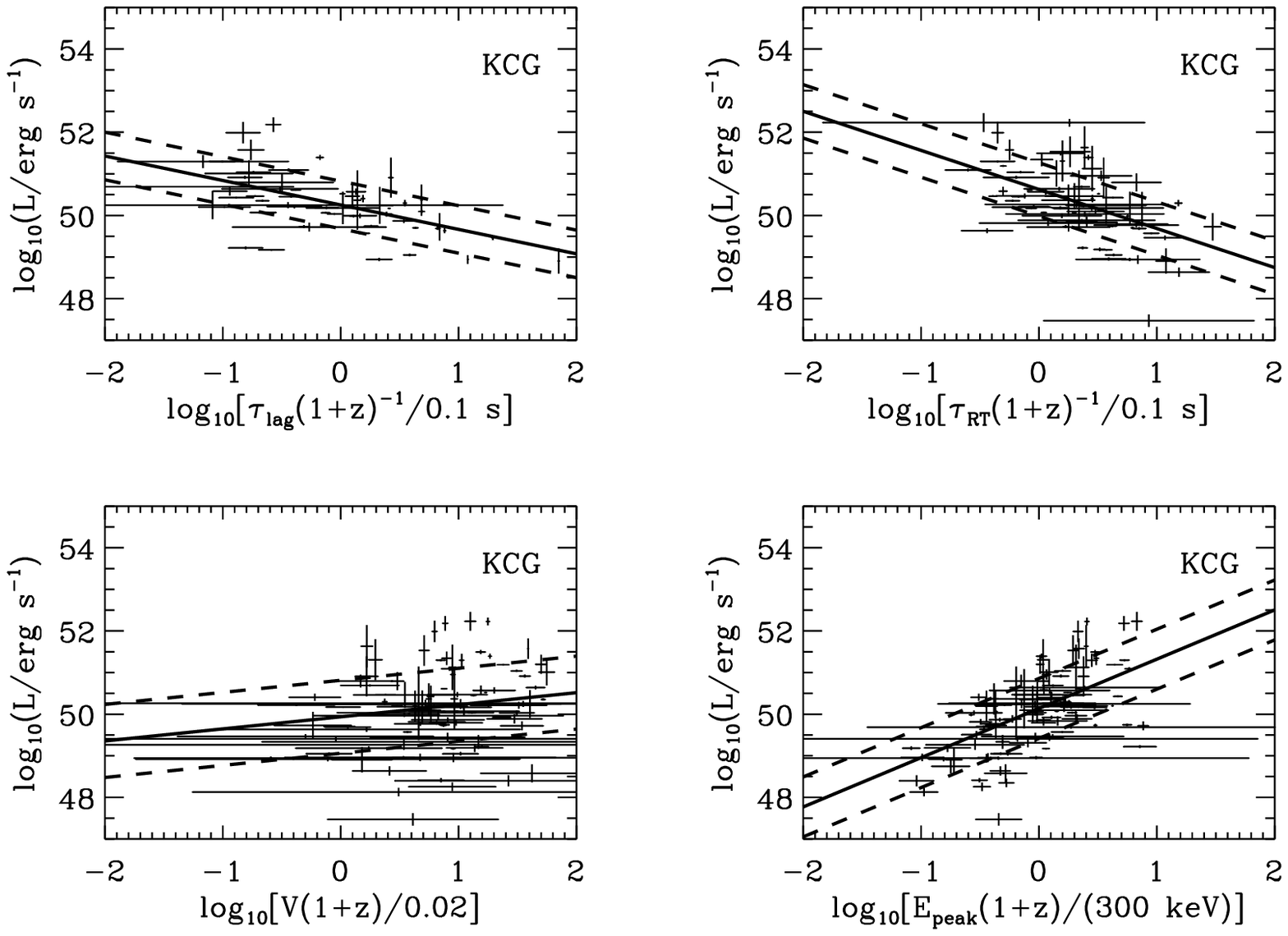}
\caption{Same as figure \ref{fig:GRBcorrLCDM} for the KCG model.}
\label{fig:GRBcorrKCG}
\end{figure*}

We consider the 115 GRBs of Xiao and Schaefer \cite{xiao09}. 
The likelihood we assume for our Bayesian analysis is reported in appendix \ref{sec:bayesParam}.
The analysis determines, at the same time, 
the cosmological parameters and the correlation coefficients.
The top panels of figures \ref{fig:GRBLCDM}, \ref{fig:GRBMann}, and \ref{fig:GRBVar} 
show the marginalized PDFs of the correlation coefficients of the relation
$L-E_{\rm peak}$ and the cosmological parameters in our three models; 
in KCG we show the parameter $\delta_0$. The PDFs of the parameters of 
the remaining correlations are similar. The parameters
of the four correlations are listed in Table \ref{tab:GRBrelations}
and the cosmological parameters are listed in Table \ref{tab:GRBcosmopars}. 
Figures \ref{fig:GRBcorrLCDM}, \ref{fig:GRBcorrCG}, and \ref{fig:GRBcorrKCG}
show the four relations in the three models. 

A visual inspection of these figures and the results in the tables show 
that all the three models have no difficulty in accommodating the current
GRB data. In addition, our results for $\Lambda$CDM are comparable to the relations shown in \cite{xiao09}. 
Our analysis also shows that, in $\Lambda$CDM, the GRBs imply 
$\Omega_0= 0.90^{+0.07}_{-0.14}$  (Table \ref{tab:GRBcosmopars}),
a factor three larger than other current measures.

\subsection{SN sample}
\label{sec:standard}

For the first time, in the section above, we have shown how the GRBs alone
can be used to extract cosmological information without
any prior information on the cosmological model and without
any support from other observables. However, current measures
of the GRB quantities are affected by large uncertainties
and the cosmological constraints are consequently rather
poor. To put more stringent constraints on the cosmological models,
we can combine the GRBs with other cosmological probes.
As an illustrative example we investigate the combination 
with SNe. Before doing so, we first apply our Bayesian analysis
to the SN data alone. 

SNe Ia are not exactly standard candles. In fact, we can not trivially
estimate their distance moduli as $\mu= m^{\rm max} - M$,
where $m^{\rm max}$ is the apparent magnitude at the peak
of the light curve in a given band and $M$ an absolute magnitude valid for all SNe,
because an empirical correlation between
colour, magnitude at peak brightness, and shape of the light curve 
exists \citep{phillips93}. 
Instead, an appropriate procedure is 
to consider, for each $i$-th SN at redshift $z_i$, 
the following empirical distance modulus in the $B$ band,
which includes corrections due to the shape-parameter $s$ of the light-curve and the color
$c=B-V(t=B_{\rm max}) + 0.057$  \citep[see, e.g.,][]{kowalski08,tripp98}   
\begin{equation}
\mu_i= m^{\rm max}_i - M +\alpha(s_i-1) -\beta c_i \; .
\label{eq:SNmu}
\end{equation}

\begin{figure*}
\includegraphics[angle=0,scale=.25]{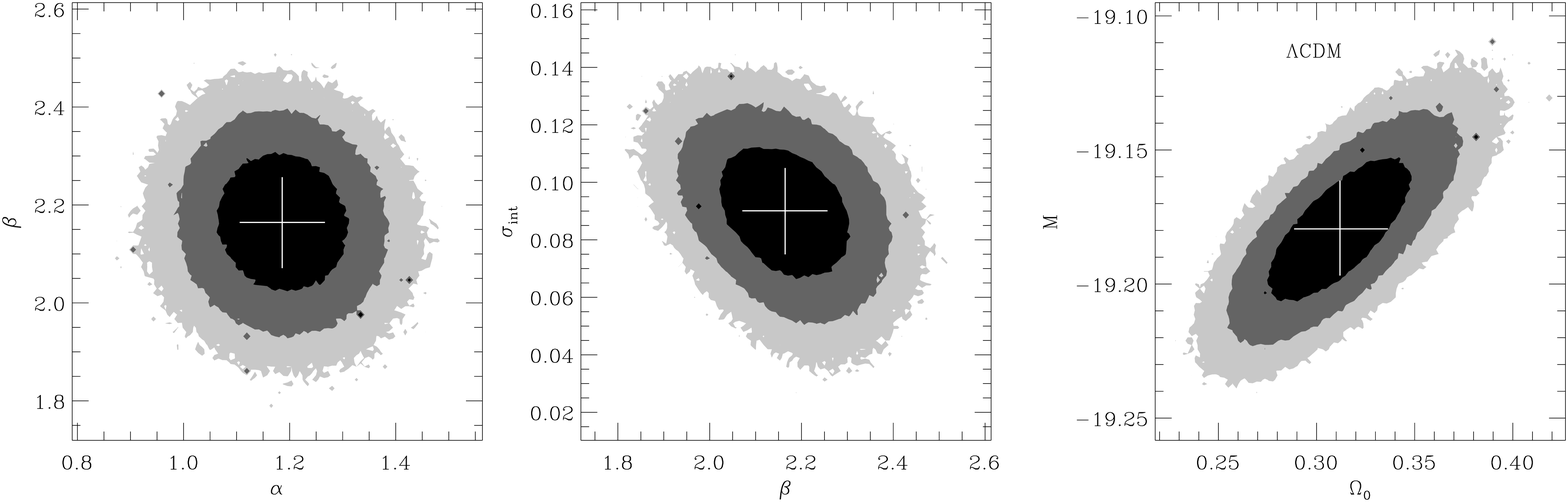}
\includegraphics[angle=0,scale=.25]{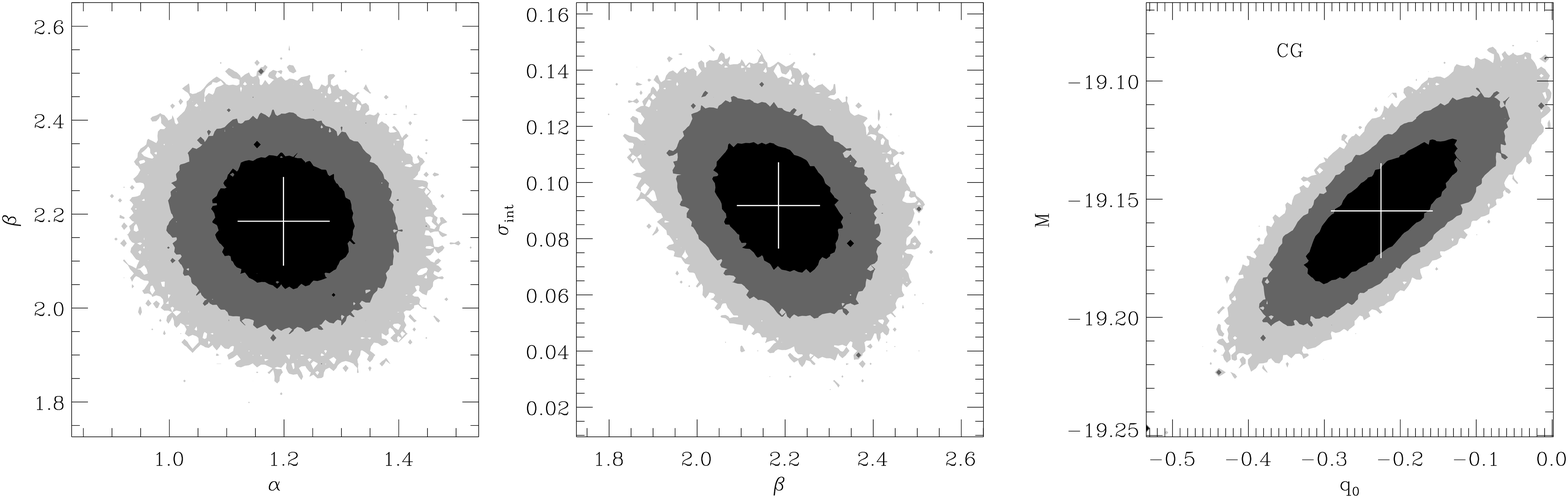}
\includegraphics[angle=0,scale=.25]{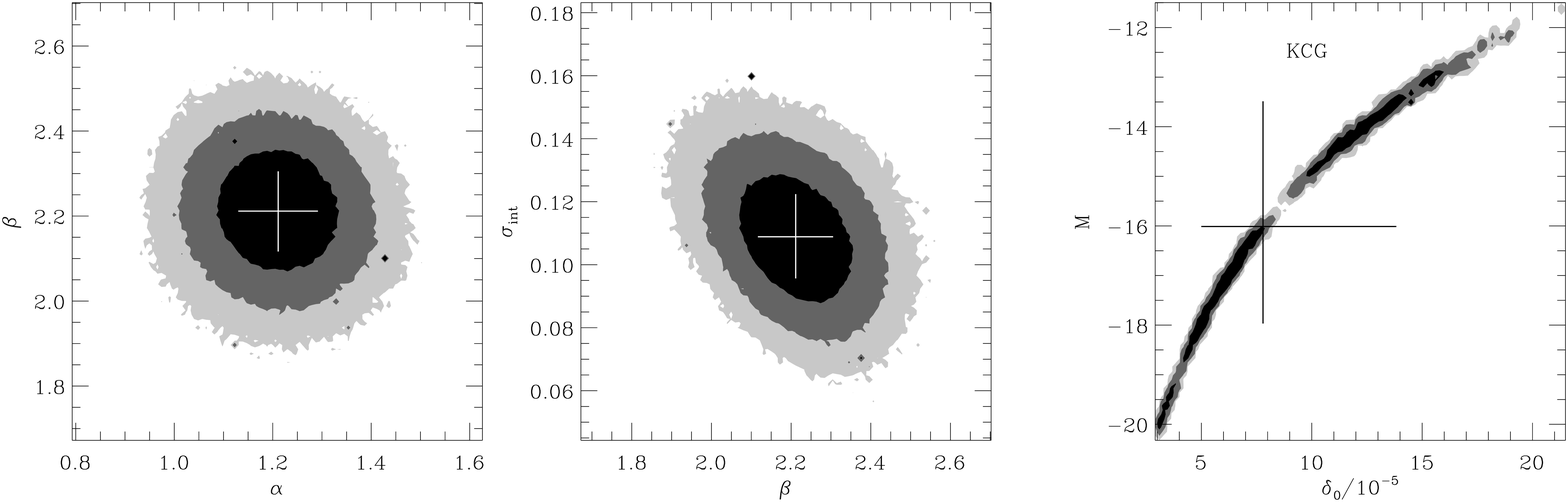}
\caption{The marginalized PDFs of the cosmological parameters of our three models
and of the SN parameters in equation (\ref{eq:SNmu})
when the Bayes analysis is applied to the SN sample alone:
$\Lambda$CDM (upper row), CG (middle row), KCG (bottom row).
Black, grey, and light-grey shaded regions 
correspond to the 68.3, 95.4, and 99.7 percent confidence levels, respectively. 
The crosses show the median values and their marginalized 1-$\sigma$ uncertainty.
}
\label{fig:ParMatrixLCDM}
\end{figure*}

\begin{figure*}
\includegraphics[angle=0,scale=.28]{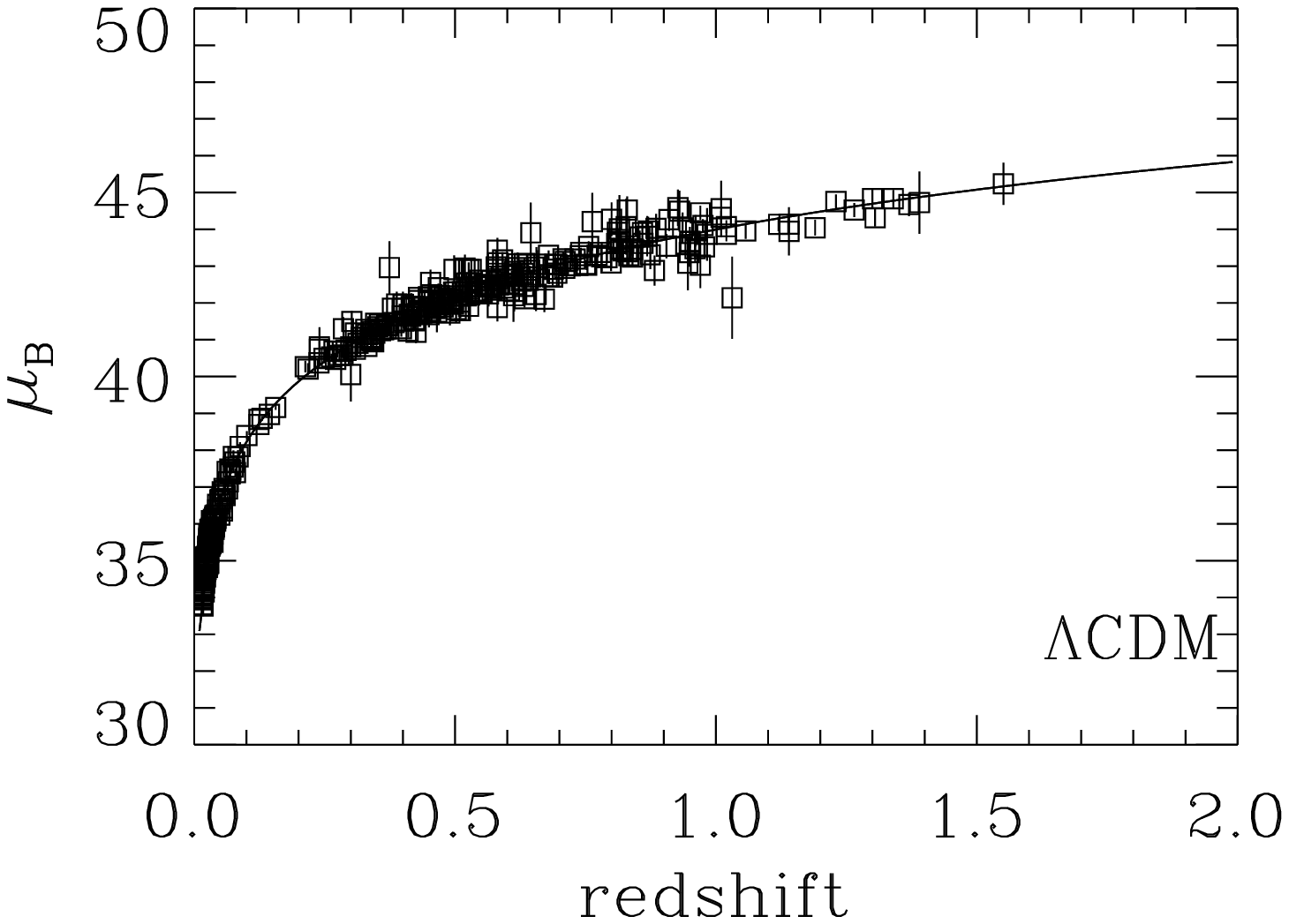}
\includegraphics[angle=0,scale=.28]{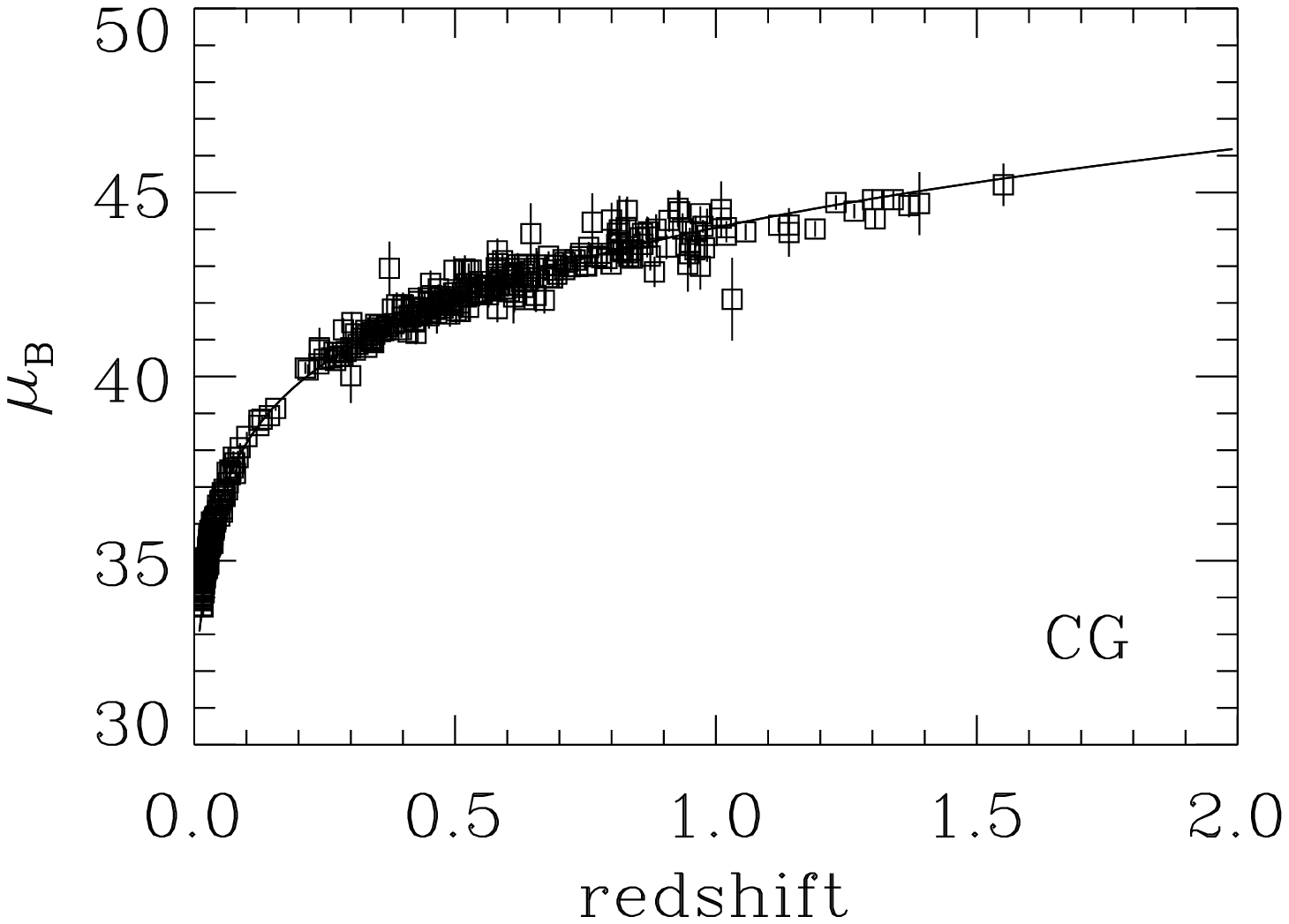}
\includegraphics[angle=0,scale=.28]{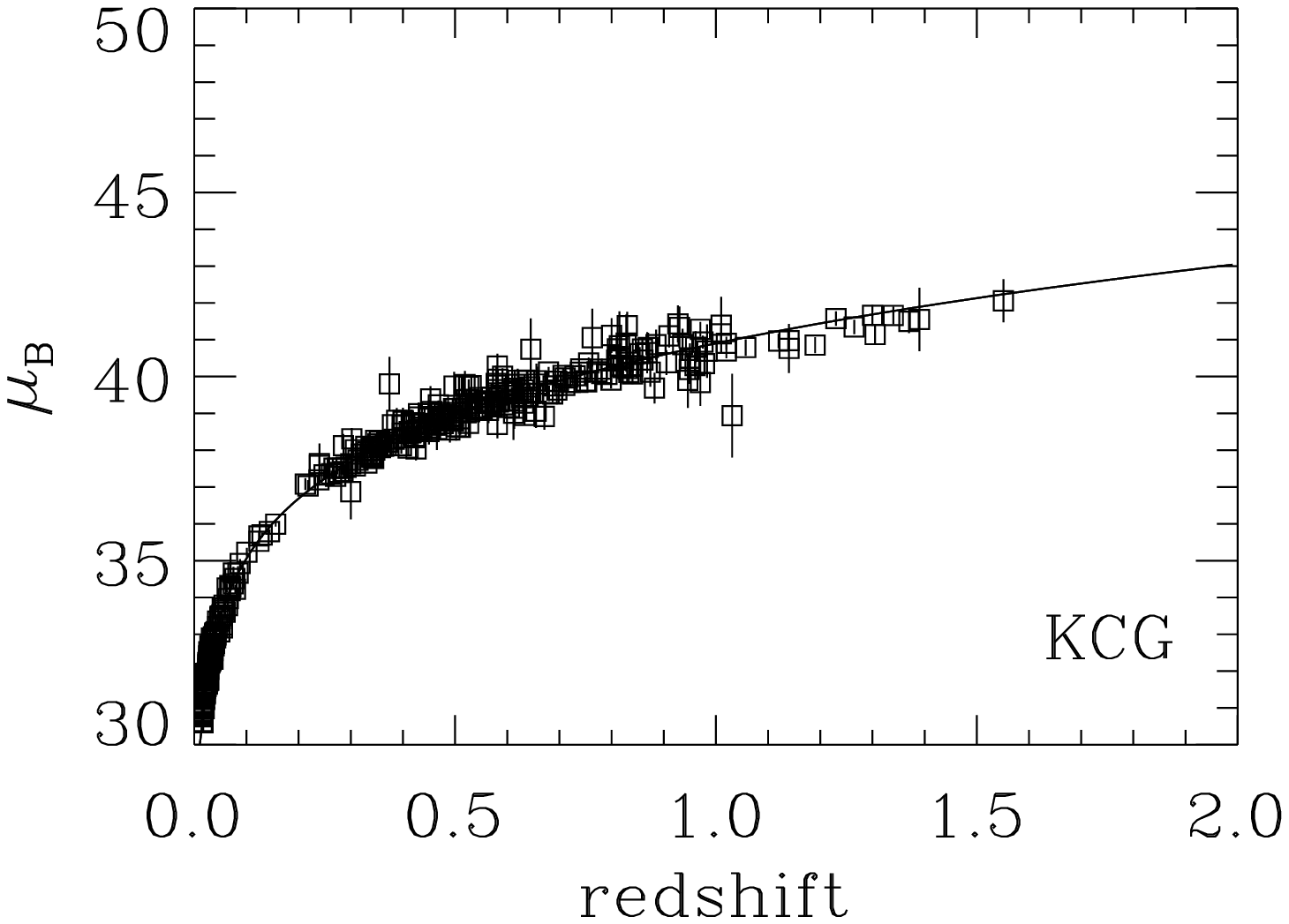}
\caption{SN Hubble diagrams derived with equation (\ref{eq:SNmu})
and the median fit parameters in our three models. The error bars are computed by considering only
the uncertainties on $m$, $s$, and $c$. The solid curves show the $\mu-z$ relation computed with
the corresponding median fit cosmological parameters.}
\label{fig:SNHD}
\end{figure*}

To estimate the three parameters $M$, $\alpha$, and $\beta$ along 
with the cosmological parameters ${\mathbf p}$, we apply the Bayesian
analysis to the data sample $\{m_i,s_i,c_i,z_i, {\mathbf S}_i\}$, 
where ${\mathbf S}_i$ is the vector of the uncertainties of the four observables
of each SN.
As in the previous section, additional hidden parameters are mimicked in the relation between the observable $m^{\rm max}$
and $\mu$, by assuming that $m^{\rm max}$ is a random variate with mean
\begin{equation}
m({\mathbf p}) = \mu({\mathbf p},z_i) + M - \alpha(s_i-1) + \beta c_i  
\end{equation}
and variance $\sigma_{\rm int}^2$.  

Whereas for $\Lambda$CDM  and CG, ${\mathbf p}=\Omega_0$ and ${\mathbf p}=q_0$,
respectively, as in the GRB analysis,
for KCG, ${\mathbf p}=(a, \delta_0)$ and the number of parameters is reduced by one.
In the $\Lambda$CDM  and CG models the expected distance modulus is 
$\mu=25+5\log (d_L/{\rm Mpc})$, where $d_L$ is given by equation (\ref{eq:lumdist})
for CG. 
In the KCG model,  $\mu$ is given instead by  
equation (\ref{eq:Varmu}).
Therefore, in this latter model the value of $H_0$ is irrelevant,
whereas for $\Lambda$CDM  and CG, $\mu$ also depends on the Hubble constant
through $d_L$. We follow the usual procedure
of removing the value of $H_0$ and its
uncertainty  by replacing $M$ with $M^\prime=M-5\log(H_0)$.

The likelihood we assume for our Bayesian analysis is reported in appendix \ref{sec:bayesParam}.
We determine, at the same time, 
the cosmological parameters, the absolute magnitude $M$ of the SNe and 
the SN parameters $\alpha$ and $\beta$, which, in turn, can
be used to compute the distance moduli of the SNe. 

We consider the 397 SNe of the Constitution set \citep{hicken09}.
Figures \ref{fig:ParMatrixLCDM}-\ref{fig:SNHD} and Table \ref{tab:SNparams}
show our results. All the three models provide the same 
pair ($\alpha, \beta$) within the confidence levels; $\Lambda$CDM and CG also provide the
same effective absolute magnitude $M$, which is fully consistent with
the absolute magnitudes
of nearby SNe \citep{riess09}. Previous
estimates of both $\Omega_0$ for $\Lambda$CDM (e.g., $\Omega_0=0.29\pm 0.01$ by \cite{march11b}) 
and $q_0$ for CG ($q_0=-0.37$ by \cite{mann03}) 
are largely within $3$-$\sigma$  
of our estimates $\Omega_0=0.312^{+0.024}_{-0.023}$ 
and $q_0=-0.225^{+0.068}_{-0.066}$, respectively. We note that the reference value of $\Omega_0$
we mention was determined by combining SNe with other probes, whereas our 
estimate of $\Omega_0$ and Mannheim's and our estimates of $q_0$
rely on SNe alone. In addition, the analysis of \cite{march11b}, unlike
ours, include the covariance matrix of each SN: based on the comparable results
we obtain, we do not expect that our approach substantially suffers from this
simplification. 

In KCG, the absolute magnitudes
of nearby SNe also appear to be within $2$-$\sigma$ of our estimate 
$M=-16.0^{+2.5}_{-2.0}$; moreover $\delta_0=3.83\times 10^{-5}$, which 
was estimated by Varieschi \cite{varieschi09b} 
by keeping $a_{\rm V}=2$ fixed, is within 
$2$-$\sigma$ of our median value $\delta_0=7.8^{+6.0}_{-2.8} \times 10^{-5}$. However, 
the parameter $M$ cannot be directly
compared to the absolute magnitudes of nearby SNe which were
estimated in a cosmological framework based on fundamentally different 
physical principles. In addition, the 68.3 percent confidence intervals are
large ($\sim 15$ percent for $M$ and $\sim 60$ percent for $\delta_0$) because $M$ and $\delta_0$ are more strongly
degenerate than $M$ and $\Omega_0$ or $q_0$ in the other models (figure \ref{fig:ParMatrixLCDM}). 
At any rate, the most important result of our analysis of KCG, that we clearly see in the
right panel of figure \ref{fig:SNHD}, is that the 
data are consistent with a SN Hubble diagram substantially
different from what it is naively assumed to be a {\it measured} Hubble diagram:
the distance moduli of SNe in this model are $~\sim$~3~mag fainter than in $\Lambda$CDM.
The reason for this result is simple:
the SN apparent magnitudes are the direct observables, whereas the SN distance moduli are not.
So, if KCG were the widely accepted standard cosmological model, and
we tested $\Lambda$CDM by assuming that the distance moduli are the direct observables, 
$\Lambda$CDM would be completely unable to describe the data. Clearly 
our approach easily prevents us from drawing this incorrect conclusion.

From figures \ref{fig:ParMatrixLCDM}-\ref{fig:SNHD} 
and Table \ref{tab:SNparams}, we conclude that the three models describe the SN data
very well. Our result shows that SNe alone do not necessarily support 
an early phase of decelerated expansion, but are consistent with 
an always accelerating universe. Amendola et al. \cite{amendola06}
show that the SN data are consistent with a universe where the
transition between the decelerated and the accelerated phase can occur as early as
$z\sim 3$. Our analysis suggests the
more extreme conclusion that the current SN data are consistent
with models where the transition phase never occurred.

\subsection{GRBs combined with SNe}

In the Bayesian approach, combining the SN and GRB samples simply
translates into considering a single sample
$S=\{{\rm SN}_i,{\rm GRB}_i\}$, 
where ${\rm SN}_i = \{m_i,s_i,c_i,z_i \}$, with corresponding uncertainties ${\mathbf S}_i$,
and ${\rm GRB}_i= \{P_{\rm bol}^i,Q_j^i,z_i\}$, with
$Q_j=\{\tau_{\rm lag}^i,\tau_{\rm RT}^i,V^i,E_{\rm peak}^i\}$
and corresponding uncertainties ${\mathbf S}_i$.
As usual, our task is the determination of the multi-dimensional probability density distribution of the
parameters $\theta=\{\alpha,\beta,M,\sigma_{\rm int}^{\rm SN}, \{a,b,\sigma_{\rm int}\}_j,{\mathbf p}\}$, 
where $ \{a,b,\sigma_{\rm int}\}_j$, $j=1,4$, are the parameters of the four GRB correlations.

The bottom panels of figures \ref{fig:GRBLCDM}, \ref{fig:GRBMann}, and \ref{fig:GRBVar}
show the marginalized PDFs of the correlation coefficients of the relation
$L-E_{\rm peak}$ and the cosmological parameters in our three models
obtained when we combine the GRB and SN samples.
Both the SN parameters (Table \ref{tab:SNparams}) and the GRB correlation parameters (Table \ref{tab:GRBrelations}) are barely affected by this combination.
On the contrary, the cosmological parameters are closer to the values obtained
with the SN sample alone than to the values obtained with the GRBs alone. 
Clearly, this is a consequence of the fact that 
the SN sample is larger and the uncertainties on the SN
measurements smaller.
As it happened with the SNe and the GRBs analysed separately, the three models describe the 
data properly.

\begin{figure*}
\includegraphics[angle=0,scale=.28]{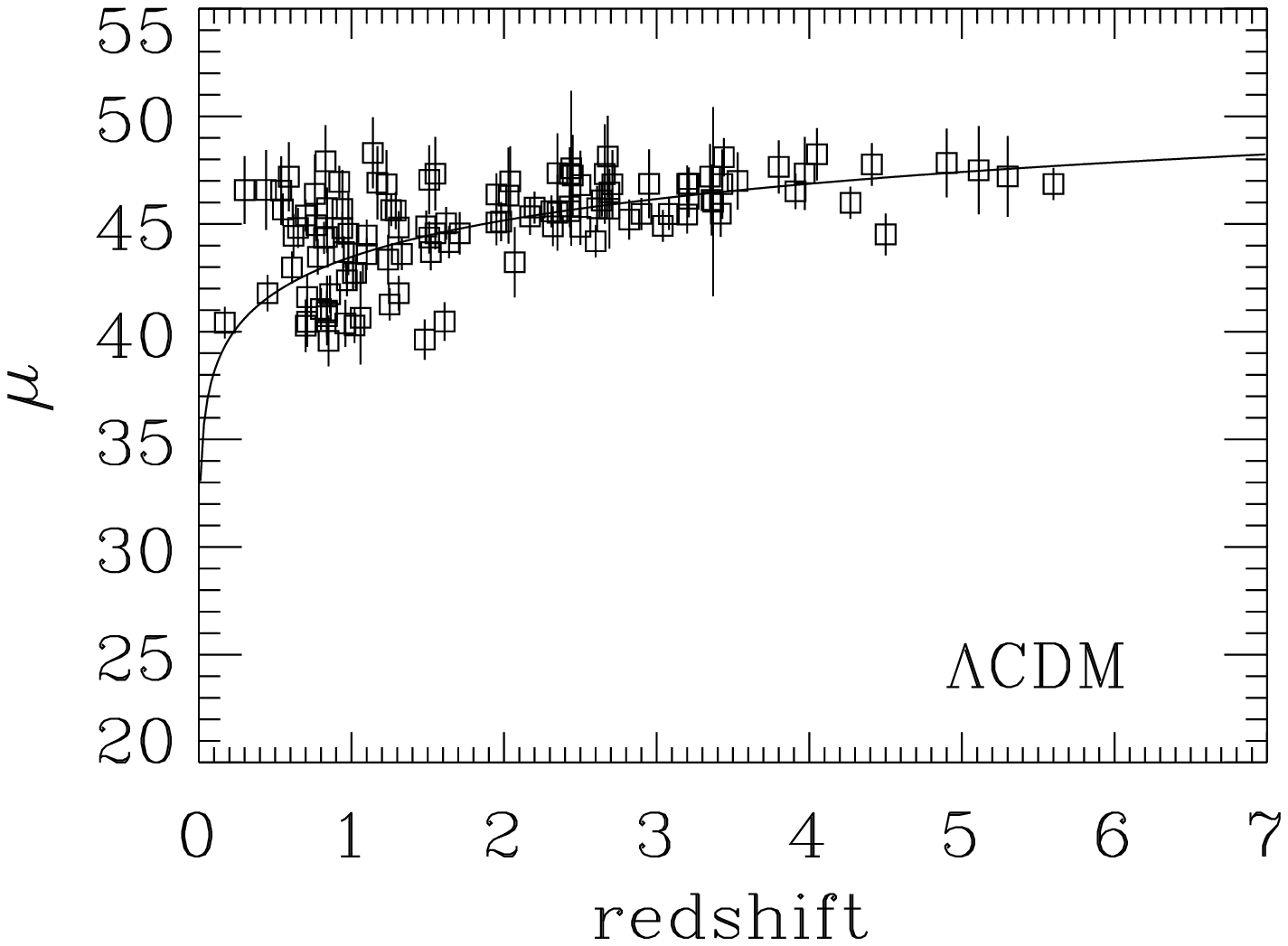}
\includegraphics[angle=0,scale=.28]{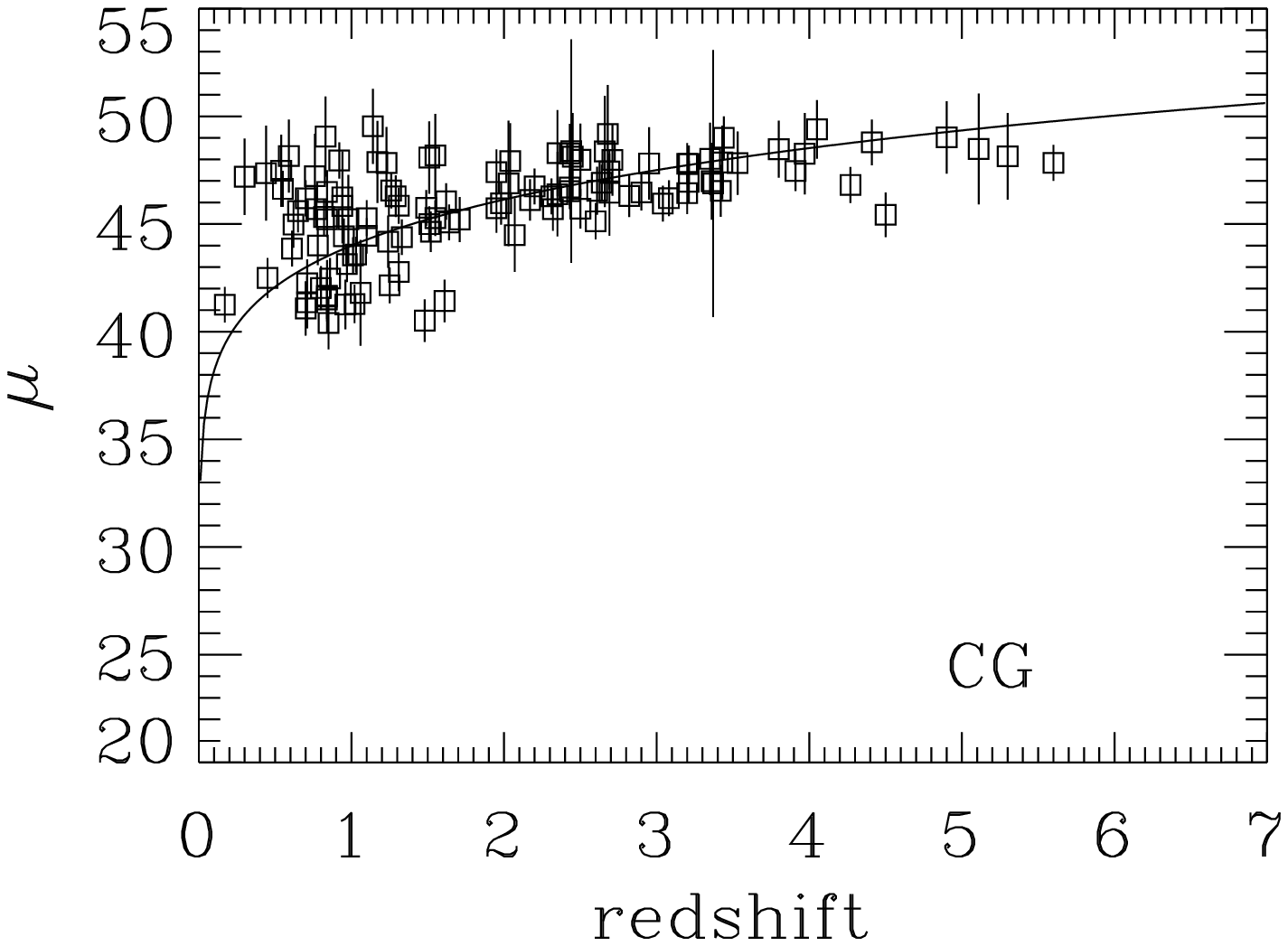}
\includegraphics[angle=0,scale=.28]{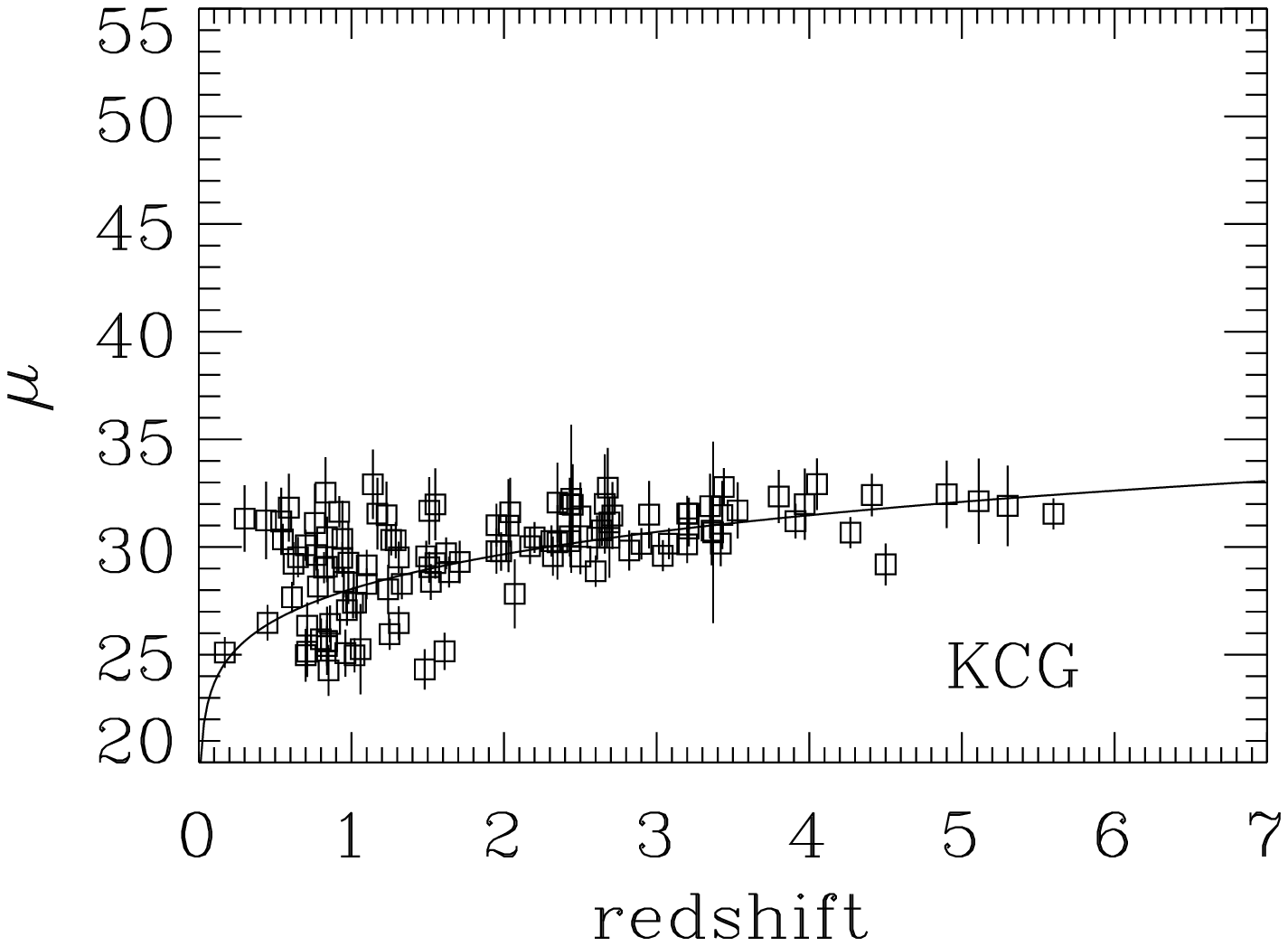}
\includegraphics[angle=0,scale=.28]{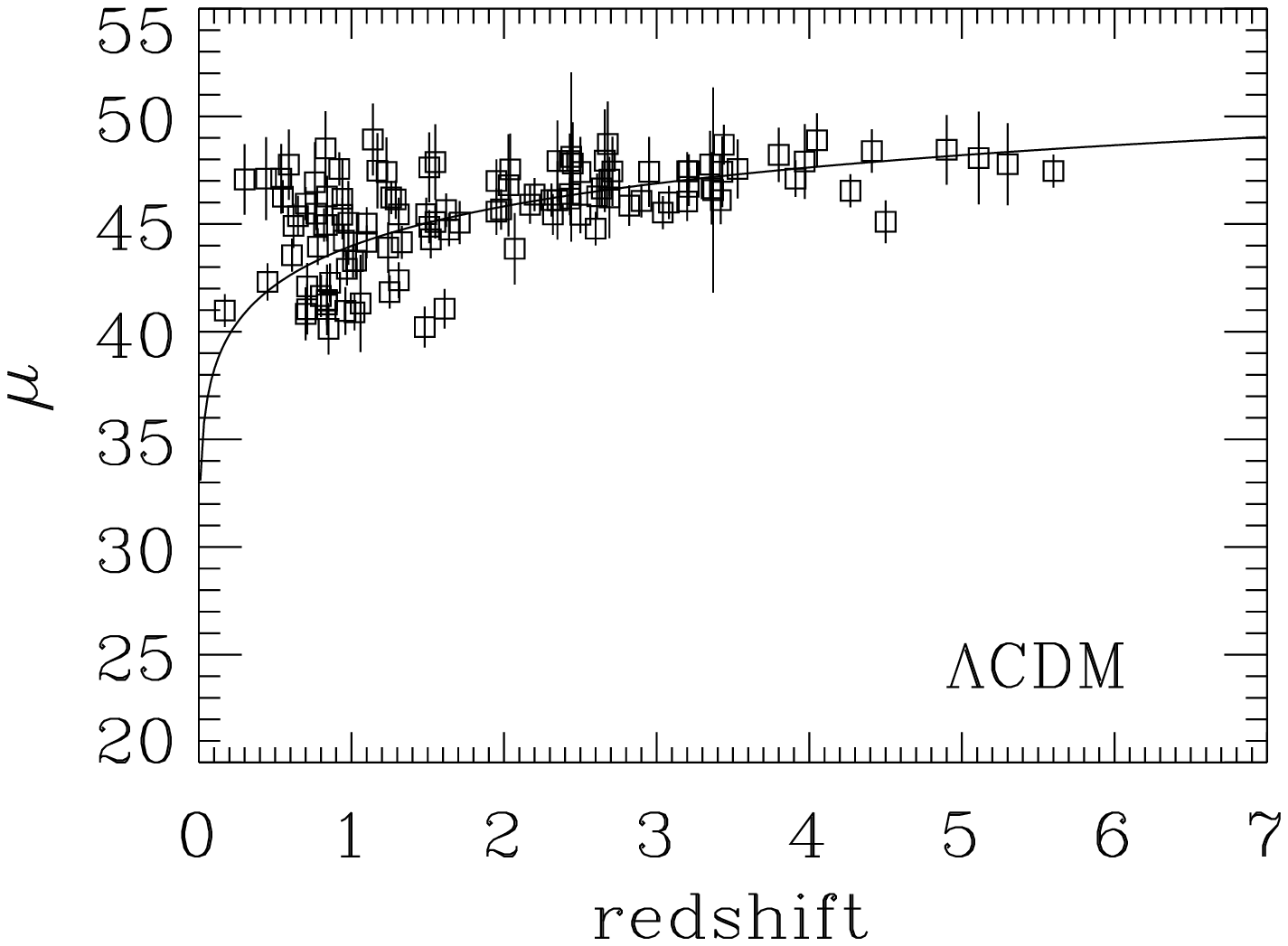}
\includegraphics[angle=0,scale=.28]{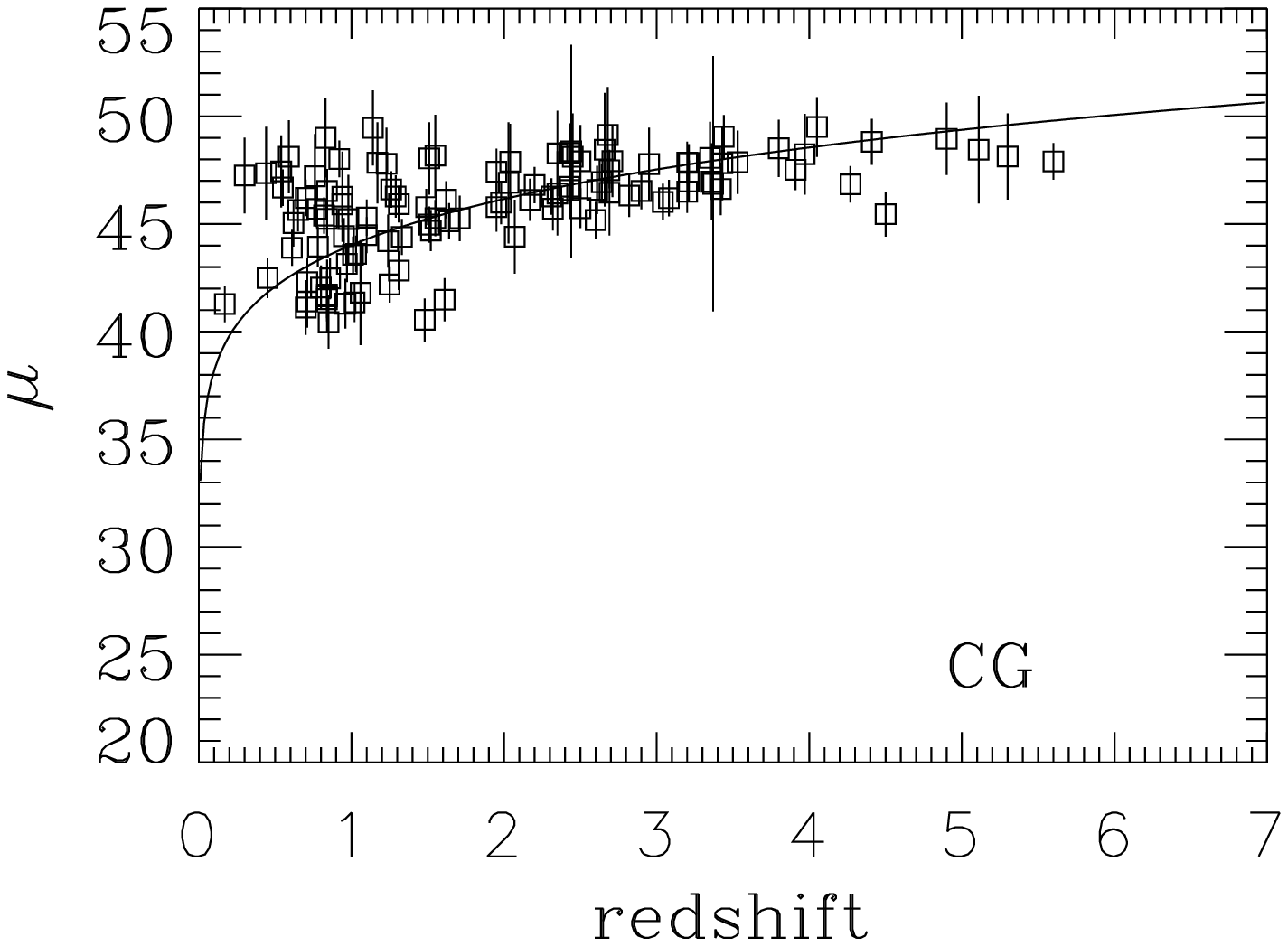}
\includegraphics[angle=0,scale=.28]{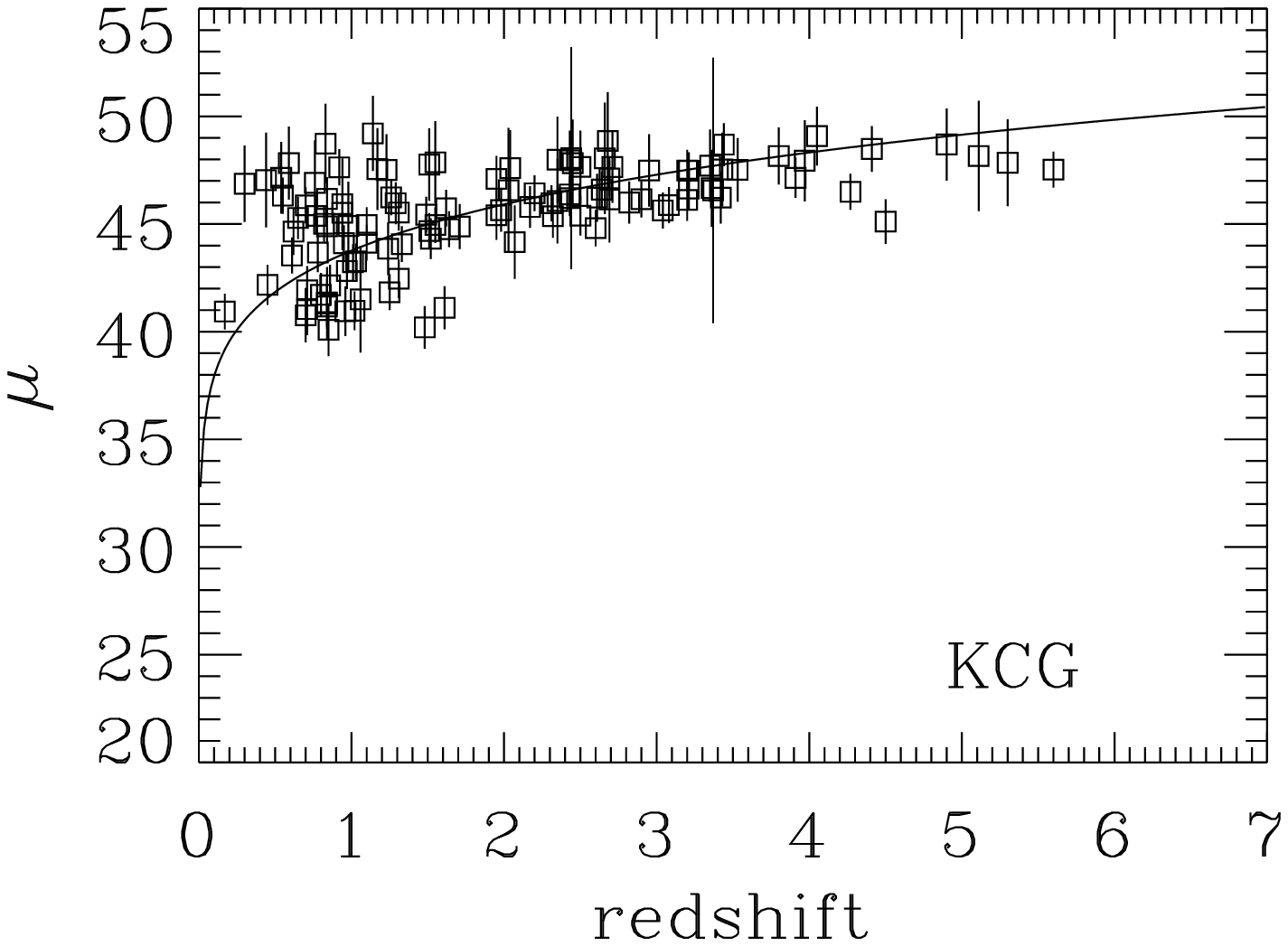}
\caption{GRB Hubble diagrams based on the GRB data alone (upper panels) and
on the sample of GRBs and SNe combined (lower panels). The solid curves are not fits 
to the data points, but the theoretical distance modulus of each 
model with the parameters derived with the Bayesian analysis.}
\label{fig:GRBHD}
\end{figure*}

We also plot the Hubble diagrams of GRBs in our three models (figure \ref{fig:GRBHD}), 
for the sake of clarity and to make our results more easily comparable to previous 
results in the literature.
To do so, we follow the standard approach \citep{schaefer07} and 
for each $i$-th GRB we compute $\mu_i^{(j)}$
where $j$ refers to one of the four relations: the directly observed quantity $Q$ provides
an estimate of $\log_{10}L=a+b\log_{10}Q$, 
which, in turns, returns $d_L$ and thus $\mu_i^{(j)}=5\log_{10}(d_L/{\rm Mpc})+25$ 
or $\mu_i^{(j)}=2.5(2+a_{\rm V})\log_{10} (d_L/d_{\rm rs})$ for the KCG model. The final distance
modulus is the weighted mean $\mu_i = \sum_{j=1}^4 w_j \mu_i^{(j)} / \sum_{j=1}^4 w_j$
where $w_j = 1/\sigma_j^2$ and $\sigma_j^2$ is the variance on $\mu_i^{(j)}$
derived with the usual error propagation law. 
Clearly, $\sigma_{\mu_i} = (\sum_{j=1}^4 w_j)^{-1/2}$
is the uncertainty on $\mu_i$. Figure \ref{fig:GRBHD} shows the Hubble diagram 
when the parameters are derived from the analysis of the GRBs alone 
(upper panels) and when the SNe are included (lower panels).
In $\Lambda$CDM and CG, the inclusion of the SNe has no visible effect.
On the contrary, in the KCG model, the effect of the SNe is dramatic: they
increase the distance moduli of GRBs by 50 percent. The solid curves are the distance moduli expected
with the cosmological parameters derived from the Bayesian analysis.  
For $\Lambda$CDM our results are similar to the results of \cite{schaefer07}. 
However, as mentioned above, both CG and KCG perform equally well, provided
that the analysis of the data is done properly.
Our result clearly shows that if we naively test alternative cosmological models
by using GRB Hubble diagrams like those shown in figure \ref{fig:GRBHD} 
as they were direct measures, we can draw grossly incorrect conclusions.

\begin{table*}
\begin{tabular}{lccccc}
\hline
\hline
relation & $N_{\rm GRB}$ & model & $a$ & $b$ & $\sigma_{\rm int}$    \\
\hline
$L-\tau_{\rm lag}$ & 59 & $\Lambda$CDM &  $51.815^{+0.094}_{-0.088}$  & $-0.59^{+0.13}_{-0.13}$   & $0.589^{+0.069}_{-0.060}$   \\
 & & &  $52.042^{+0.085}_{-0.091}$   & $-0.61^{+0.14}_{-0.14}$    & $0.609^{+0.078}_{-0.060}$      \\
 & & CG & $52.157^{+0.099}_{-0.095}$   & $-0.66^{+0.15}_{-0.14}$  & $0.661^{+0.078}_{-0.064}$  \\
 & & & $52.170^{+0.091}_{-0.099}$  &  $-0.68^{+0.15}_{-0.18}$   & $0.67^{+0.10}_{-0.07}$   \\
 & & KCG &  $50.3^{+0.7}_{-1.1}$  & $-0.59^{+0.14}_{-0.15}$  & $0.571^{+0.073}_{-0.063}$    \\
 & & & $55.28^{+0.44}_{-0.85}$  & $-0.66^{+0.14}_{-0.14}$ & $0.664^{+0.075}_{-0.066}$  \\
\\
$L-\tau_{\rm RT}$ & 79 &$\Lambda$CDM &  $52.12^{+0.10}_{-0.10}$  &  $-0.98^{+0.15}_{-0.16}$  & $0.640^{+0.068}_{-0.057}$   \\
&  & &  $52.44^{+0.10}_{-0.10}$  &  $-1.03^{+0.16}_{-0.17}$   & $0.651^{+0.072}_{-0.061}$     \\
 & & CG & $52.59^{+0.11}_{-0.11}$   & $-1.13^{+0.17}_{-0.17}$   &  $0.678^{+0.071}_{-0.061}$ \\
 & & & $52.60^{+0.10}_{-0.10}$     & $-1.12^{+0.18}_{-0.18}$ &  $0.684^{+0.081}_{-0.066}$  \\
 & & KCG & $50.6^{+0.7}_{-1.1}$  & $-0.94^{+0.18}_{-0.17}$  & $0.643^{+0.079}_{-0.060}$   \\
 & & & $55.72^{+0.43}_{-0.84}$ & $-1.14^{+0.17}_{-0.18}$  & $0.677^{+0.070}_{-0.061}$   \\
\\
$L-V$ & 104 & $\Lambda$CDM  & $51.50^{+0.16}_{-0.15}$ &  $0.31^{+0.17}_{-0.17}$  & $0.894^{+0.076}_{-0.069}$    \\
 &  &  & $51.70^{+0.16}_{-0.16}$  &  $0.34^{+0.19}_{-0.17}$   & $0.908^{+0.079}_{-0.074}$      \\
 & & CG & $51.74^{+0.18}_{-0.17}$  & $0.47^{+0.19}_{-0.19}$  & $0.939^{+0.083}_{-0.074}$     \\
 & &  & $51.75^{+0.16}_{-0.17}$    & $0.44^{+0.20}_{-0.16}$  & $0.941^{+0.073}_{-0.075}$  \\
 & & KCG & $49.9^{+0.7}_{-1.1}$  & $0.29^{+0.18}_{-0.18}$ & $0.877^{+0.079}_{-0.069}$  \\
 & & & $54.87^{+0.45}_{-0.86}$   & $0.46^{+0.19}_{-0.20}$   & $0.937^{+0.087}_{-0.072}$  \\
\\
$L-E_{\rm peak}$ & 115 &$\Lambda$CDM & $51.694^{+0.081}_{-0.079}$    & $1.20^{+0.17}_{-0.17}$  & $0.721^{+0.059}_{-0.051}$    \\
 & & & $51.917^{+0.080}_{-0.080}$    & $1.22^{+0.18}_{-0.17}$ & $0.735^{+0.065}_{-0.055}$   \\
 & & CG & $52.055^{+0.087}_{-0.080}$    &  $1.32^{+0.19}_{-0.18}$   & $0.778^{+0.063}_{-0.057}$    \\
 & & & $52.051^{+0.086}_{-0.095}$  & $1.31^{+0.20}_{-0.21}$  & $0.773^{+0.062}_{-0.064}$    \\
 & & KCG & $50.1^{+0.7}_{-1.1}$  & $1.18^{+0.14}_{-0.17}$ &  $0.720^{+0.080}_{-0.055}$  \\
 & & & $55.19^{+0.43}_{-0.83}$  & $1.32^{+0.18}_{-0.18}$  & $0.777^{+0.062}_{-0.055}$  \\
\hline
\end{tabular}
\caption{Median fit parameters of the GRB relations obtained from the analysis of the GRBs alone (first row
of each model) and from the analysis of the GRBs and SNe combined (second row of each model). 
The uncertainties are the marginalized 68.3 percent confidence intervals. The second column lists
the number of GRBs used for the estimate of each correlation.}
\label{tab:GRBrelations}
\end{table*}

\begin{table*}
\begin{tabular}{lcc}
\hline
\hline
  model & ${\mathbf p}$ &   \\
\hline
  $\Lambda$CDM & $\Omega_0$ & $0.90^{+0.07}_{-0.14}$   \\
               &           & $0.324^{+0.026}_{-0.025}$  \\
\\
 CG &  $q_0$ & $-0.12^{+0.08}_{-0.16}$  \\
     & & $-0.164^{+0.015}_{-0.022}$   \\
\\
 KCG &  $\delta_0/10^{-5},\quad a_{\rm V},\quad a({\mathbf t_0})$  &  $17^{+9}_{-11},\quad  1.035^{+0.066}_{-0.025},\quad 5.5^{+3.3}_{-4.0}$ \\
      & &  $4.0^{+1.1}_{-0.7},\quad  2.038^{+0.013}_{-0.013},\quad 5.3^{+3.0}_{-3.3}$  \\
\hline
\end{tabular}
\caption{Median fit cosmological parameters of the GRB relations obtained from the analysis of the GRBs alone (first row
of each model) and from the analysis of the GRBs and SNe combined (second row of each model).
The uncertainties are the marginalized 68.3 percent confidence intervals.}
\label{tab:GRBcosmopars}
\end{table*}

\begin{table*}
\begin{tabular}{lcccccc}
\hline
\hline
model & $\alpha$ & $\beta$ & $M$ &  $\sigma_{\rm int}$ &  ${\mathbf p}$ &     \\
\hline
$\Lambda$CDM &  $1.186^{+0.081}_{-0.080}$ &  $2.164^{+0.092}_{-0.093}$  & $-19.179^{+0.018}_{-0.017}$  & 
$0.090^{+0.015}_{-0.015}$   & $\Omega_0$ & $0.312^{+0.024}_{-0.023}$    \\  
 &  $1.182^{+0.088}_{-0.082}$  & $2.153^{+0.097}_{-0.092}$ & $-19.172^{+0.019}_{-0.018}$ & $0.0917^{+0.017}_{-0.016}$ &  & $0.324^{+0.026}_{-0.025}$   \\  
\\
CG &  $1.199^{+0.081}_{-0.080}$ & $2.185^{+0.094}_{-0.095}$  &  $-19.155^{+0.020}_{-0.020}$ & $0.0918^{+0.015}_{-0.015}$  & $q_0$ & $-0.225^{+0.068}_{-0.066}$  \\ 
 &  $1.193^{+0.098}_{-0.091}$  & $2.18^{+0.12}_{-0.10}$   & $-19.142^{+0.014}_{-0.017}$ & $0.092^{+0.016}_{-0.017}$  & & $-0.164^{+0.015}_{-0.022}$    \\ 
\\
KCG &  $      1.211^{+    0.081}_{-0.081}$  & $2.212^{+0.094}_{-0.095}$   &  $-16.0^{+2.5}_{-2.0}$  & $     0.109^{+0.014}_{-0.013}$  &  $\delta_0/10^{-5},\quad a_{\rm V}$ & $7.8^{+6.0}_{-2.8},\quad 2.044^{+0.013}_{-0.013}$    \\      
 &  $1.204^{+0.082}_{-0.084}$ & $2.21^{+0.10}_{-0.10}$  &  $-18.9^{+1.1}_{-0.9}$ & $0.110^{+0.014}_{-0.013}$ & & $4.0^{+1.1}_{-0.7},\quad 2.038^{+0.013}_{-0.013}$    \\      
\hline
\end{tabular}
\caption{Median fit parameters of the SN data alone (first row of each model)
and when combined with the GRBs (second row of each model). The uncertainties are the marginalized 68.3 percent confidence intervals.}
\label{tab:SNparams}
\end{table*}

\begin{table*}
\begin{tabular}{lcc}
\hline
\hline
  sample & $\Lambda$CDM/CG & $\Lambda$CDM/KCG    \\
\hline
  GRBs &  37.9 & 12.0    \\
   SNe &  6.6  & 7.2  \\
  GRBs+SNe & 1.5 & 24.3  \\
\hline
\end{tabular}
\caption{Values of $\ln B_{12}$, where $B_{12}$ is the Bayes
factor in equation (\ref{eq:Bfactor}): $B_{12}>1$ favours model $M_1$ over model $M_2$.}
\label{tab:BayesFactor}
\end{table*}

\section{Discussion}
\label{sec:realdisc}

The Bayesian analysis described above shows that current
data from SNe and GRBs can be easily described by both
$\Lambda$CDM, that requires a deceleration/acceleration
transition in the expansion of the Universe, and conformal
gravity, that requires an always accelerating universe.

To understand quantitatively which model the GRB and SN data might favour, we
compute the Bayes factor $B_{12}$ defined in equation (\ref{eq:Bfactor}). 
We assume flat priors for all the parameters of the models but the internal
dispersions, as explained in Appendix \ref{sec:bayesParam}. When 
we consider the GRB sample alone, we find
that $\Lambda$CDM is favoured over CG by a factor $\ln B_{12}=37.9$,
and over KCG by a factor $\ln B_{12}=12.0$ (Table \ref{tab:BayesFactor}).
When we consider the SN sample alone, our Bayes factor estimate shows that
$\Lambda$CDM is favoured over CG and KCG by a
Bayes factor $\ln B_{12}=6.6 $ and $\ln B_{12}=7.2 $, respectively. 
Finally, for the SN and GRB sample combined, the Bayes factor again shows that
$\Lambda$CDM is favoured, by a factor $\ln B_{12}=1.5$ over CG and by a factor $\ln B_{12}=24.3$
over KCG.
According to Jeffreys' table \cite{jeffreys61},\footnote{Jeffreys \cite{jeffreys61} separates the
values of the Bayes factor $B_{12}$ into the six intervals
$[(<1),(1,3),(3,10),(10,30),(30,100),(>100)]$, where larger $B_{12}$ values favour model
$M_1$ over model $M_2$ more strongly. It became customary to indicate
the intervals in natural logarithms $[(<0),(1,1.1),(1.1,2.3),(2.3,3.4),(3.4,4.6),(>4.6)]$.
The choice of these intervals clearly
is arbitrary and other, albeit comparable, choices were also adopted \citep[e.g.,][]{trotta07}.}
in most cases these values are large and $\Lambda$CDM would be ``decisively'' favoured over CG and KCG.

However, we emphasize that this conclusion is not as robust as it
might appear. There is in fact an extended literature on how the Bayes factor is sensitive to the 
choice of the priors \citep[e.g.,][]{aitkin91, kass95, weakliem99, berger01, johnson05, liu08}. Investigating this issue is beyond the scope
of this work, but it is clear that different prior choices
might substantially change our Bayes factor estimates and, consequently, the conclusion they suggest. 

Overall, our analysis shows that GRBs are substantially unable to distinguish between two cosmological models
that have very distinct predictions on the early expansion history of
the Universe, where the GRBs are expected to provide unambiguous results. 
This failure is mostly due to the very large uncertainties on the GRB observables.
In fact, our results obtained by combining GRBs with SNe confirm that
the cosmological constraints are actually driven by the 
SNe rather than the GRBs \citep{li08}.
Furthermore, GRB afterglow energies corrected for beaming span two
orders of magnitude \citep{lil08, mcbreen10},
and current detectors can introduce
significant bias against hard bursts in GRB samples \citep{shah09}.
It thus seems difficult
to consider GRBs as standard candles, unless we are able to correct for these systematic effects,
which appear to be substantial.
We necessarily conclude that GRBs do not represent effective cosmological probes.
Currently, enlarging the SN sample at $z>1$ might remain a more promising method to 
constrain the expansion history of the Universe.

Additional tests of conformal gravity might come
from the Bayesian analysis of the combination of different cosmological
probes deriving from both the expansion history of
the Universe and the properties of cosmic structures.
CG seems to have difficulties in describing the thermodynamics of
clusters of galaxies \citep{horne06,diaf09}, whereas no investigation of
structure formation in KCG is available yet.
However, conformal gravity remains worth investigating for the elegant
solution it suggests to the cosmological constant, zero-point
energy and quantum gravity problems without resorting to
either dark matter or dark energy \citep{mann10, mann11}.
 
The additional merit of the models that largely depart
from $\Lambda$CDM, like CG and KCG, is that they provide 
predictions that are clearly distinct from the standard model
expectations. Our results teach us 
that we do need a proper technique to test the data
against different models: using a method that partly relies
on the models we want to test does not generally yield sensible
results, unless the models are very similar to each other. In 
this case however, the conclusions we draw are weak.
Using a really model-independent technique returns results
that are definitely more robust. 
For example, we show that KCG, where the SN distance moduli
are $\sim$~3~mag fainter than in $\Lambda$CDM,
also can describe the data. We thus confirm that with the
Bayesian technique we can robustly and self-consistently 
test various cosmological models, no matter how different 
they are.

\section{Conclusion }
\label{sec:disc}

Current data from SNe and, more recently, from GRBs are usually interpreted
as supporting evidence for the $\Lambda$CDM model, where the 
expansion history of the Universe has an early deceleration
phase, down to $z\sim 1$, followed by the acceleration phase that lasts to the present time.
Based on this supposedly
robust interpretation of the available data, most alternative cosmological models were conceived
to reproduce these two distinct phases. On the
contrary, conformal gravity was proposed
well before the observation of high-$z$ SNe and it predicts a Universe
expansion that has been always accelerating. This model is thus ideal
to test whether the data undoubtly support $\Lambda$CDM.
  
We perform our test with two  variants of conformal gravity: CG and KCG. 
KCG is an even more drastic alternative model than CG, because a number of physical
quantities, including the flux from distant sources and the cosmological
redshift, have a radically different interpretation than in the standard model.
Therefore, unlike many cosmological tests described in the literature, 
ours requires a method that does not rely on any prior information from the same model 
that has to be tested. 

We lay out the problem within a full Bayesian context, and 
perform three different analyses of a sample of GRBs, a sample
of SNe and the sample of the GRBs and SNe combined. 
With our Bayesian approach, we simultaneously estimate 
the PDFs of the parameters of the
GRB correlations, the parameters of the SN distance moduli and the cosmological parameters.

Contrary to the expectation, we show that the current data can
be described by CG, KCG, and $\Lambda$CDM equally well.
The application of our method to the SN sample alone shows 
that the data even support the SN distance moduli derived
in KCG, where their physical interpretation is fundamentally different from
$\Lambda$CDM, and are $\sim 3$~mag fainter than in the standard
model. Similarly, the cosmological 
information that can be derived from the GRBs does also support CG and KCG
when the information is extracted properly.

We conclude that, at face value, the $\Lambda$CDM expansion 
history is not supported by the data as robustly as it is naively believed,
but other wildly different models can still describe the data satisfactorily.
Therefore, current data are unable to exclude that
the Universe has been always accelerating: 
both variants of conformal gravity we have investigated here are, from 
the point of view of the background expansion history of
the Universe, viable alternatives to $\Lambda$CDM.

$\Lambda$CDM is favoured over CG and KCG only when we
resort to the Bayes factor computed by assuming flat priors on the model parameters. 
It will be crucial to investigate whether $\Lambda$CDM 
still remains favoured for different choices of the priors or, more interestingly, 
when  we include additional cosmological probes 
based on the formation of large-scale structure. In this context, $\Lambda$CDM requires
large amounts of dark matter that is expected to be unnecessary in conformal gravity.

\acknowledgments
We sincerely thank Johannes Buchner and Michael Gruberbauer for developing their superb
code APEMoST and making it available to the community
({\tt apemost.sourceforge.net}). 
We also are particularly grateful to Johannes Buchner
and Stefano Andreon for intense and enlightening correspondence on Bayesian statistics. 
Stefano Andreon is also acknowledged for a very stimulating seminar, delivered in Torino,
on Bayesian statistics applied to astrophysics that inspired this work.
We finally thank Margaret Geller, Xiao-Li Meng and the JCAP Editor, Carl Akerlof, 
for valuable suggestions on the presentation of our results.
Support from the INFN grant PD51 and the PRIN-MIUR-2008 grant \verb"2008NR3EBK_003"  
``Matter-antimatter asymmetry, 
dark matter and dark energy in the LHC era'' is gratefully acknowledged.
LO also acknowledges support from a 2009 National 
Fellowship ``L’OR\'EAL Italia Per le Donne e la Scienza''
 of the L’OR\'EAL-UNESCO program “For Women in Science” 
and partial support from the ASI Contract No. I/016/07/0 COFIS.
This research has made use of NASA's Astrophysics Data System.

\appendix
\section{Bayesian analysis}
\subsection{Bayesian parameter estimation}\label{sec:bayesParam}

Consider a model $M$ described by a set of parameters $\theta$ with
probability $p(\theta\vert M)$ of occurring.
The probability of measuring the set of data $D$
when the model $M$ is described by the parameters 
$\theta$ is the likelihood $p(D\vert \theta,M)$. 
The probability of observing a set of data $D$ is thus
\begin{equation}
p(D\vert M) =  \int p(D\vert \theta,M) p(\theta\vert M) {\rm d}\theta \; ;
\end{equation}
$p(D\vert M)$ is called the Bayesian evidence of the model $M$,
$p(\theta\vert M) $ the prior.
We are interested in estimating the PDF of the parameters given our data set $D$
\begin{equation}
p(\theta\vert D, M)={ p(D\vert\theta, M)p(\theta\vert M) \over p(D\vert M)} \; . 
\end{equation}
For this task, we need to assume a likelihood $p(D\vert \theta,M)$.

For the GRBs, $D= \{P_{\rm bol}^i,\{Q^i_j\}_{j=1,4},z^i, {\mathbf S}^i\}$,
$\{Q^i_j\}_{j=1,4}=\{\tau_{\rm lag}^i,\tau_{\rm RT}^i,V^i,E_{\rm peak}^i\}$, 
$ {\mathbf S}^i$ is the vector of the uncertainties of $\{P_{\rm bol}^i,\{Q^i_j\}_{j=1,4}\}$,
$\theta=\{\{a_j,b_j,\sigma_{{\rm int}_j}\}_{j=1,4},{\mathbf p}\}$, ${\mathbf p}$ is
the vector of the cosmological parameters, 
and we assume the likelihood
\begin{equation}
p(D\vert\theta,M) = \prod_{j=1}^4\prod_i {1\over (2\pi \sigma_{ij}^2)^{1/2}} \exp\left[-(P_{\rm bol}^i-W^i_j)^2 \over 2\sigma_{ij}^2\right]  
\end{equation}
where
\begin{equation}
\sigma_{ij}^2=\sigma_{{\rm int}_j}^2 + \sigma_{P_{\rm bol}^i}^2 + b_j^2 \sigma_{Q^i_j}^2
\end{equation}
and
\begin{equation}
 W^i_j = a_j + b_j Q^i_j - f(z^i;{\mathbf p})\, \quad j=1,\dots,4 \;  
\end{equation}
where $f(z^i;{\mathbf p})$ is the proper function of the luminosity distance $d_L$.
$W^i_j$ is the mean of the random variate $\log_{10}P_{\rm bol}^i$, whose
variance is $\sigma_{{\rm int}_j}^2$, according to the $j$-th correlation.

For the SNe, $D=\{m_i,s_i,c_i,z_i, {\mathbf S}_i\}$, where ${\mathbf S}_i$ is
the vector of the observable uncertainties. If 
$\mu(z_i; {\mathbf p})$ is the distance modulus, and $m_i$ a random variate with mean 
\begin{equation}
W_i = \mu(z_i; {\mathbf p}) + M - \alpha(s_i-1) + \beta c_i 
\end{equation}
and variance $\sigma_{\rm int}^2$, the complete
set of parameters is $\theta=\{M,\alpha,\beta,\sigma_{\rm int},{\mathbf p}\}$.
We assume the likelihood 
\begin{equation}
p(D\vert\theta,M) = \prod_i {1\over (2\pi \sigma_i^2)^{1/2}} \exp\left[-(m_i-W_i)^2 \over 2\sigma_i^2\right] 
\end{equation}
where
\begin{equation}
\sigma_i^2=\sigma_{\rm int}^2 + \sigma_{m_i}^2 + 
\alpha^2\sigma_{s_i}^2 + \beta^2\sigma_{c_i}^2 + \sigma_{\mu_i}^2\; ; 
\end{equation}
$\sigma_{m_i}$, $\sigma_{s_i}$, $\sigma_{c_i}$ are the uncertainties of the 
observables and $\sigma_{\mu_i}$ derives from the uncertainty $\sigma_{z_i}$ on 
the SN redshift $z_i$.

For both the SN and GRB samples, we assume independent flat priors for all the $\theta$ 
parameters except for the internal dispersions $\sigma_{\rm int}$, which are
positive defined. In this case we assume
\begin{equation}
p(\sigma_{\rm int}\vert M) = {\mu^r\over \Gamma(r)}  x^{r-1} \exp(-\mu x) 
\end{equation}
where $x=1/\sigma_{\rm int}^2$, and $\Gamma(r)$ is the usual gamma function.
This PDF describes a variate with mean $ r/\mu$, and variance $ r/\mu^2$.
We set $r=\mu=10^{-5}$ to assure an almost flat prior.

\subsection{Bayesian model selection}\label{sec:bayesEvidence}

The probability of the model $M$ to be correct, given the set of data $D$ is, according
to Bayes' theorem, the model posterior probability
\begin{equation}
p(M\vert D)= {p(D\vert M) p(M)\over p(D)} \; .
\end{equation}
When comparing two models $M_1$ and $M_2$, we can compute the ratio of the posterior
probabilities
\begin{equation}
{p(M_1\vert D)\over p(M_2\vert D)} = B_{12} {p(M_1)\over p(M_2)}
\end{equation}
where
\begin{equation}
B_{12} = {p(D\vert M_1)\over p(D\vert M_2)}
\label{eq:Bfactor}
\end{equation}
is the Bayes factor. If $p(M_1)=p(M_2)$, 
$B_{12}>1$ clearly favours model $M_1$; $M_2$ is favoured otherwise.
Estimating the Bayesian evidence $p(D\vert M)$ of a model thus provides a tool to compare different
models for a given set of data, if the models are equally probable.

The computation of the Bayesian evidence is not a trivial task \citep[see, e.g.,][for a review]{trotta08}.
We use here the thermodynamic integration (or parallel tempering; \cite[e.g.,][]{hobson03}).
The posterior probability for the model parameters is
\begin{equation}
p(\theta\vert D,M) = {p(D\vert \theta,M) p(\theta\vert M)\over p(D\vert M)} \; .
\end{equation}
The evidence $p(D\vert M)$ is a normalization constant; we can thus define the
unnormalized probability density 
\begin{equation}
q(\theta\vert D,M) = p(D\vert \theta,M) p(\theta\vert M) \; .
\end{equation}
Consider the variant 
\begin{equation}
q_\beta(\theta\vert D,M) = [p(D\vert \theta,M)]^\beta p(\theta\vert M)
\end{equation}
and
\begin{equation}
p_\beta(\theta\vert D,M) = {[p(D\vert \theta,M)]^\beta p(\theta\vert M)\over p_\beta(D\vert M) } 
\end{equation}
with the parameter $\beta\in[0,1]$. Because $\int p_\beta(\theta\vert D,M)  {\rm d}\theta =1$,
the evidence is 
\begin{eqnarray}
 p_\beta(D\vert M) & = & \int[p(D\vert \theta,M)]^\beta p(\theta\vert M) {\rm d}\theta \cr
 & = & \int q_\beta(\theta\vert D,M) {\rm d}\theta  \; .
\end{eqnarray}
Taking the derivative $\partial \ln  p_\beta(D\vert M)/\partial\beta = [1/p_\beta(D\vert M)]
\partial  p_\beta(D\vert M)/\partial\beta$ yields
\begin{eqnarray}
{\partial \ln  p_\beta(D\vert M)\over \partial\beta}& = &{\int \ln p(D\vert\theta, M) q_\beta(\theta\vert D,M) {\rm d}\theta \over \int q_\beta(\theta\vert D,M) {\rm d}\theta} \cr
& = & \int \ln p(D\vert\theta, M) p_\beta(\theta\vert D,M){\rm d}\theta \cr
& = & \langle \ln p(D\vert\theta, M) \rangle_\beta 
\end{eqnarray}
where the mean is intended over the set of parameters $\theta$ and with respect to the posterior
probability for the model parameters $p_\beta(\theta\vert D,M)$.
The integral $\int_0^1 [\partial \ln  p_\beta(D\vert M)/ \partial\beta] {\rm d}\beta = \ln p_1(D\vert M) - \ln p_0(D\vert M)=
 \ln p(D\vert M)$ (because $\int p(\theta\vert M){\rm d}\theta=1$) yields
the logarithm of the Bayesian evidence entering the Bayes factor in equation (\ref{eq:Bfactor}) 
\begin{equation}
  \ln p(D\vert M) = \int_0^1  \langle \ln p(D\vert\theta, M) \rangle_\beta {\rm d}\beta\; .
\end{equation}

\subsection{Numerical details}\label{sec:numdet}

For our Bayesian analysis we use the code APEMoST developed by Johannes Buchner and 
Michael Gruberbauer \citep{buchner11}.
The first version of the code was applied to 
the analysis of stellar pulsations \citep{gruberbauer09}.

We use $2\times 10^6$ MCMC iterations 
to guarantee a fairly complete sampling of the parameter space and twenty chains or values
of $\beta\in[0,1]$. The boundaries of the parameter space were set to
$[-1000,1000]$ for all the $a$ and $b$ parameters, $[0.01,1000]$ for the $\sigma_{\rm int}$ parameters, 
$[0.01,1.0]$ for $\Omega_0$, $[-1,-0.001]$ for $q_0$, $[1,3]$ for $a_{\rm V}$, $[3,30]\times 10^{-5}$
for $\delta_0$, and $[10^{-5},10]$ for $a({\mathbf t}_0)$. 
The initial seed of the random number generator was set
with the {\tt bash} command {\tt \verb"GSL_RANDOM_SEED=$RANDOM"}.
 
\bibliography{references}
\bibliographystyle{JHEP}

\end{document}